\documentclass[showpacs,,showkeys,preprintnumbers,amsmath,amssymb]{revtex4}
\usepackage{amsmath}
\usepackage{dcolumn}
\usepackage{bm}
\usepackage{graphicx}
\usepackage{subfigure}

\newcommand{\N}{\mathbb{N}}
\newcommand{\Z}{\mathbb{Z}}
\newcommand{\R}{\mathbb{R}}

\newcommand{\ke}[1]{| #1 \rangle}

\newcommand{\bk}[2]{\langle #1 | #2 \rangle}

\newtheorem{thm}{Theorem}
 
\begin{document}

\title{Isochronism and tangent bifurcation of band edge
  modes in Hamiltonian lattices}

\author{J. Dorignac${}^1$, S. Flach${}^2$}

\affiliation{${}^1$ College of Engineering, Boston University, 44 Cummington
  St., Boston,
  Massachussets 02215.\\ 
${}^2$ Max-Planck-Institut f\"ur Physik komplexer Systeme,
N\"othnitzer
Stra\ss e 38, D-01187 Dresden}

\date{\today}

\begin{abstract}
In {\em Physica D} 
{\bf 91}, 223 (1996) \cite{Fla96}, results were obtained regarding the
tangent bifurcation of the band edge modes ($q=0,\pi$) 
of nonlinear Hamiltonian lattices made of $N$ coupled oscillators. 
Introducing the concept of 
{\em partial isochronism} which characterises the way the frequency 
of a mode, $\omega$, depends on its energy, $\varepsilon$, we generalize
these results and  
show how the bifurcation energies of these 
modes are intimately connected to their degree of isochronism. 
In particular we prove that in a lattice of coupled purely isochronous
oscillators ($\omega(\varepsilon)$ strictly constant), 
the in-phase mode ($q=0$) never undergoes a tangent bifurcation
whereas the out-of-phase mode ($q=\pi$) does, provided the strength of the
nonlinearity in the coupling is sufficient. We derive a discrete 
nonlinear Schr\"odinger equation governing the slow modulations of 
small-amplitude band edge modes and show that its nonlinear exponent
is proportional to the degree of isochronism of the corresponding orbits. 
This equation may be seen as a link between the tangent bifurcation of band 
edge modes and the possible emergence of localized modes such as discrete 
breathers.

\end{abstract}

\maketitle

\vskip2pc

\section{Introduction} \label{Intro}

Since almost fifteen years now, properties of discrete breathers (DB) 
in nonlinear translationally invariant 
Hamiltonian lattices are under intense investigation. 
General features regarding these time
periodic, spatially localised excitations are 
well understood and have been documented in several reviews
\cite{Sievers95,Aubry97,Flachrep98,MacKay00,FlachZol01,Dauxois04,Campbell04}. 
Mathematical proofs of their existence go back to a paper by R.S. MacKay and
S. Aubry \cite{MacKay94} which considers lattices of interacting
oscillators. It shows the possibility to continue 
single site oscillations in the decoupled (so-called
anti-continuous) limit to nonzero coupling between the oscillators
provided the corresponding orbit stays out of resonance with the 
low amplitude lattice modes. 
This, in particular, requires the oscillator (onsite) potential to be non
isochronous, that is, to possess orbits whose frequency 
varies with the energy. Other works, either generalising
this method \cite{Sepulchre97}, or using different approaches 
\cite{Flach95,James01,Aubry01} have added to the variety of 
rigorous existence proofs of discrete breathers. 

One of us has performed an analytical study 
of the way band edge modes (BEMs) may bifurcate (tangently) to give rise
to new periodic orbits breaking the translation invariance of the lattice
\cite{Fla96}. It
is a common conjecture indeed that discrete breathers are among the orbits
bifurcating from these plane waves. The bifurcation analysis
investigates the possible existence of almost extended discrete breathers
and, in this respect, is complementary to MacKay's and Aubry's theorem which
proves the existence of strongly peaked ones in networks of 
weakly coupled non isochronous oscillators. 
It provides the critical energy at 
which tangent bifurcations of BEMs possibly occur according to the mode under
consideration (see also \cite{bb83}). But it is restricted to the generic case
where plane wave orbits do not bear any degree of isochronism.  

In order to discuss further the specific properties of
partially isochronous BEMs, we now define this concept more precisely.
Let us first remark that in one-dimensional (1D) convex 
potentials, the motion is always 
periodic. To any given energy corresponds
a unique orbit whose frequency is determined by the features of the 
potential (basically, its shape). 
We will say an orbit to be {\em isochronous} 
if its frequency does not depend on its energy. 
Given the one-to-one correspondence between
the potential and its orbits, if the motion is isochronous 
the potential can be said to be isochronous as well. 

In higher dimension however, the
concept of isochronous potentials becomes ambiguous due to the 
possible presence of several families of periodic orbits, each of 
which having its own energy-frequency dependence.   
{\em Isochronism is then a property of a particular family of orbits} rather
than a property of the potential itself. Let us notice though, 
that to give birth to a family of isochronous orbits, 
the potential must fulfill certain conditions. These
constraints simply vary according to the motion under consideration, and may
be different for different families of periodic orbits of one and the same
potential. 

The most famous example of a 1D isochronous potential is the harmonic well
$V(x)=\omega^2 x^2/2$ whose frequency $\omega$ is well known to be  
energy-independent. 
Nevertheless, isochronism is not the privilege of the latter and it can be
shown that appropriate shears of the parabolic curve produce other 
{\em non symmetric} isochronous potentials $V(x) \neq
V(-x)$ \cite{Bolotin}. 

Now, for generic convex potentials, the
frequency of a given periodic orbit 
can be expanded at low energies $E$ (bottom of the potential) 
as a power series in $E$.
Its behaviour is generally linear with $E$ around the 
equilibrium position. We will call "partially isochronous" or, more
precisely {\em isochronous up to order} $n$, orbits whose
frequency behaves instead as a nonlinear function of the energy 
when expanded around $E=0$. 
Typically, $\omega^2(E) = \omega_0^2 +\gamma_n E^n + \circ
(E)^n$, $n \geq 2$, $\omega_0 > 0$ and $\gamma_n \neq 0$. 
For $n=1$, we recover the case of non isochronous
motions or, equivalently in our terminology, of orbits
isochronous up to order 1. Completely isochronous orbits verify
$\omega^2(E) = \omega_0^2$.     

The aim of the present
paper is to extend the results \cite{Fla96} to the class of potentials 
rendering certain periodic motions partially or completely isochronous. 
As we will see, it turns out that the tangent bifurcation of band edge plane
waves is intimately related to the low-energy behaviour of their frequency
(hence the above definitions). 
This remark will enable us to generalise the perturbative analysis performed
in \cite{Fla96} and to express the bifurcation energy of BEMs in a very simple
way valid for any potential. An immediate conclusion of this generalisation
will be that, {\em families of discrete breathers bifurcating from partially
isochronous BEMs possess energy thresholds, even in 1D}. 
This, to some extent, shall be confirmed by a ``multiple-scale'' analysis which
leads to a discrete nonlinear Schr\"odinger equation (DNLS)
whose nonlinear exponent is related to the degree of isochronism of the BEM.
The existence of energy thresholds of DBs for
such DNLS models has been confirmed \cite{Flach97}  and
has been proved by M.I. Weinstein
\cite{Weinstein99}.  
         
The paper is organised as follows: in section \ref{Sec2}, we present a 
low-energy perturbative solution of the in-phase
and out-of-phase modes and in section \ref{Sec3}, we use them to derive the
constraints imposed to 1D or 2D potentials to enforce a partial isochronism of
these motions. In section \ref{Sec4}, we study the linear stability of these
orbits and derive an expression for the energy at which they undergo a tangent
bifurcation. Finally, in
section \ref{Sec5}, we discuss the implications of these results and
present a brief multiple-scale analysis which corroborates our conclusions.

\section{Equations of motion} \label{Sec2}
\subsection{Generalities}
All along this paper, we will investigate the dynamical properties of a lattice
described by the following Hamiltonian
\begin{equation} \label{Ham}
 H = \sum_{n=1}^N \left[ \frac{1}{2} p_n^2 + V(x_n) + W(x_{n+1}-x_n) \right].
\end{equation}
with periodic boundary conditions $x_{n+N} = x_n$. For the sake of
simplicity, we consider an even number of sites $N$ ranging from 2 to
infinity.
The onsite ($V(x)$) and the interaction ($W(x)$) potentials are both assumed to
possess a minimum at $x=0$ around which they can be expanded as
\begin{equation} \label{expandVW}
V(x) = \sum_{\mu = 2}^{\infty} \frac{1}{\mu} v_{\mu} x^{\mu} \ \ \ ; \ \ \ 
W(x) = \sum_{\mu = 2}^{\infty} \frac{1}{\mu} \phi_{\mu} x^{\mu}. 
\end{equation}
The first coefficients of these expansions, $v_2$ and $\phi_2$,
represent harmonic frequencies and are assumed to be strictly positive. 
The Hamiltonian equations of motion for (\ref{Ham}) are given by
\begin{eqnarray}
 \dot{x}_n &=& p_n \, , \nonumber \\
\dot{p}_n &=& - V'(x_n) - W'(x_{n}-x_{n-1}) + W'(x_{n+1}-x_{n}). \label{lattx} 
\end{eqnarray}
Let us introduce the normal coordinates
\begin{equation}
 Q_q = \frac{1}{N} \sum_{n=1}^N e^{iqn} x_n \, , \ \ \ q = \frac{2\pi l}{N}, 
\ l \in \{-\frac{N}{2}+1,..,\frac{N}{2} \}. \label{Qofx}
\end{equation} 
Their properties are
\begin{equation}
 Q_{q+2\pi} = Q_q \ \ \ \text{and} \ \ \ Q_{-q} = Q^*_q \ \ (x_n \in \R), 
\end{equation}
and inverting the transform (\ref{Qofx}) yields
\begin{equation}
 x_n = \sum_q e^{-iqn} Q_q . \label{xofQ}
\end{equation} 
Rewritten in terms of normal coordinates, equations (\ref{lattx})
read now
\begin{equation}\label{eqmotQ}
\ddot{Q}_q + F_q (Q) = 0,
\end{equation}
where
\begin{equation}
F_q (Q) = \frac{1}{N} \sum_{n=1}^{N} e^{iqn} \left[
V'\left(\sum_{q'} e^{-iq'n} Q_{q'}\right)+
W'\left(\sum_{q'} (1-e^{iq'}) e^{-iq'n} Q_{q'}\right)-
W'\left(\sum_{q'} (e^{-iq'}-1) e^{-iq'n} Q_{q'}\right) \right]. \label{Fvec}
\end{equation}
A linearization of $F_q(Q)$ around $Q_q = 0$ leads to the equations of
motion of a harmonic lattice, namely
\begin{equation}\label{eqmotQlin}
\ddot{Q}_q + \omega^2_{q,0} Q_q = 0,
\end{equation}
where 
\begin{equation} \label{omega2q0}
\omega^2_{q,0} = v_2 +4 \phi_2 \sin^2 \left( \frac{q}{2}\right)
\end{equation}
represents the squared frequency of each {\em linear} mode $q$ (hence the
additional subscript 0, the frequency of the nonlinear mode being
denoted by $\omega_q$). 

In what follows, we will be interested in the stability of two
particular nonlinear modes corresponding to the natural continuation
of the linear $q=0$ and $q=\pi$ modes defined by
(\ref{eqmotQlin}). These nonlinear modes are periodic solutions of
(\ref{eqmotQ}) which converge to their respective linear modes as
their energy tends to zero. Notice that the linear frequency
$\omega_{0,0}$ of the in-phase mode is always nondegenerate
and because the number of sites $N$ is even, the linear frequency
$\omega_{\pi,0}$ of the out-of-phase mode is nondegenerate as well. All
other modes $q \neq 0,\pi$ are twofold degenerate ($\omega_{q,0} =
\omega_{-q,0}$).
In the next two sections, we define more precisely the two nonlinear
in-phase and out-of-phase modes and evaluate them perturbatively by means of a
Poincar\'e-Lindstedt expansion carried out at low energy.  

\subsection{In-phase mode (orbit I)} \label{subsecinp}
\subsubsection{Equation of motion}
Oscillators are said to be in phase when they perform identical
periodic motions. This corresponds to
\begin{equation} 
Q_q = Q_0 \, \delta_{q,0} \label{OrbI}
\end{equation} 
where $ \delta_{q,q'}=1$ if $q=q' \ [2\pi]$ and 0 else. 
The previous expression is a solution of the equations of
motion (\ref{eqmotQ}) provided
\begin{equation}
\ddot{Q}_0 + V'(Q_0) = 0 \, .\label{inp}
\end{equation} 
The solution $Q_q = Q_0 \delta_{q,0}$ represents the in-phase periodic
orbit. We call it orbit I. 
\noindent
The total energy of the lattice evolving according to orbit I is
\begin{equation}
E_I = H(\{ x_n=Q_0\}) = N \left(\frac{1}{2} \dot{Q}_0^2+V(Q_0) \right)\, .
\label{Einp}
\end{equation}
We will use an energy density (or energy per site) rather than the
total energy $E_I$ to describe this orbit. It is given by 
$\varepsilon_I = E_I/N$ and represents the energy of the oscillator
$Q_0$ evolving according to (\ref{inp}).

\subsubsection{Solution at low energy} \label{sssIIB2}

Eq. (\ref{inp}) represents the motion of a single oscillator in the
potential $V(x)$. According to our assumptions regarding the latter,
the potential is convex in $x=0$ and at small energy, the motion
is bounded and thus periodic. We can solve for it in perturbation by
expanding the solution as a Poincar\'e-Lindstedt series (see
e.g. \cite{Nayfey}).
For this purpose, let us first define a new dimensionless time $\tau = \omega
(\varepsilon) t$, where $\omega (\varepsilon)$ is the frequency of
$Q_0$ as a function of its energy $\varepsilon$ (for the sake of
clarity we have dropped the subscript I). The corresponding period is
$T(\varepsilon)=2\pi/\omega(\varepsilon)$. We show in Appendix
\ref{App1} how this period can be expanded as a power series in
energy. Let us also define $X=(2\varepsilon/v_2)^{1/2}$, as well as
the new dimensionless quantities $\tilde{T}(X)=T(\varepsilon)\sqrt{v_2}/(2\pi)$  
and $\tilde{V}(x)=V(x)/v_2$.
Equation (\ref{inp}) now reads 
\begin{equation} \label{dimlessinp}
\frac{\partial^2 Q_0(X,\tau)}{\partial \tau^2} + \tilde{T}^2(X)
\tilde{V}'\left(Q_0(X,\tau)\right)=0, 
\end{equation}    
where we have explicitly mentioned the energy dependence of $Q_0$ on $X$.
To solve this last equation, we expand $Q_0(X,\tau)$ as a series in $X$ around
0:
\begin{equation} \label{expandQ0inX}
Q_0(X,\tau) = \sum_{n=1}^{\infty} Q_0^{(n)}(\tau) X^n \, .
\end{equation}   
We choose without lost of generality an initial condition such that
$\dot{Q}_0(X,0) = 0$ which means that $Q_0(X,0)$ is a turning point of the
potential $V$ defined by the relation $V(Q_0(X,0))=\varepsilon \Leftrightarrow
\tilde{V}(Q_0(X,0))=X^2/2$. Inverting the previous relation gives
\begin{equation} \label{Q0tau0}
Q_0(X,0) = X + \sum_{n=2}^{\infty} \sigma_n X^n \, ,
\end{equation}
where the first odd coefficients $\sigma_n$ are given in appendix \ref{App1}.
According to the relation above, 
the initial conditions for the functions $Q_0^{(n)}(\tau)$ are
\begin{equation} \label{Q0tau0ini}
Q_0^{(1)}(0) = 1\ \ \ ; \ \ \ Q_0^{(n)}(0) = \sigma_n \, , \ \forall n \geq
2\, \ \ \ \text{and}\ \ \ \dot{Q}_0^{(n)}(0) = 0 \, , \ \forall n \geq 1 .
\end{equation}
Moreover, according to (\ref{per3}), we know the explicit form of
$\tilde{T}(X)$ which reads,
\begin{equation} \label{TtildeofX}
\tilde{T}(X) = 1 + \sum_{k=1}^{\infty} \tilde{T}_{2k} X^{2k}\ \ \ \text{where}\ \ \ 
\tilde{T}_{2k} = \sigma_{2k+1} \frac{(2k+1)!!}{(2k)!!} \, .
\end{equation}
Reinstating (\ref{TtildeofX}) and (\ref{expandQ0inX}) in (\ref{dimlessinp})
and expanding as a series of $X$, we derive a set of differential equations for
$Q_0^{(n)}(\tau)$. This way is clearly related to the
Poincar\'e-Lindstedt method except that the preliminary calculation \eqref{TtildeofX} of the
period $\tilde{T}(X)$ automatically removes all secular terms
from equation (\ref{dimlessinp}). 

Using the results of Appendix \ref{App1}, we may derive for example
the first differential equations involving
$Q_0^{(1)}(\tau),Q_0^{(2)}(\tau)$ and $Q_0^{(3)}(\tau)$.
\begin{eqnarray}
  & & \ddot{Q}_0^{(1)}(\tau)+Q_0^{(1)}(\tau)=0 \, , \\
& & \ddot{Q}_0^{(2)}(\tau)+Q_0^{(2)}(\tau)+\alpha_2 \left[
  Q_0^{(1)}(\tau)\right]^2 =0 \, , \\
& &
\ddot{Q}_0^{(3)}(\tau)+Q_0^{(3)}(\tau)+\left(\frac{5}{6}\alpha_2^2-\frac{3}{4}\alpha_3\right)Q_0^{(1)}(\tau)
+\alpha_3 \left[Q_0^{(1)}(\tau)\right]^3 + 2\alpha_2 Q_0^{(1)}(\tau) Q_0^{(2)}(\tau)=0 \, , 
\end{eqnarray}
where the double dot stands now for a differentiation with respect to
$\tau$ and $\alpha_n =v_{n+1}/v_2$. Solving this system together with the initial conditions (\ref{Q0tau0ini}), we obtain
\begin{eqnarray} \label{Q0123inp}
 Q_0^{(1)}(\tau) & = &  \cos(\tau) \, , \\
Q_0^{(2)}(\tau) & = & \frac{1}{6}\alpha_2 \left(\cos(2\tau)-3\right)\, , \\
Q_0^{(3)}(\tau) & = &
\frac{1}{96}\left(2\alpha_2^2+3\alpha_3\right)\cos(3\tau)+
\left(\frac{37}{144}\alpha_2^2-\frac{9}{32}\alpha_3\right)\cos(\tau)\, .   
\end{eqnarray}
where we have used $\sigma_2 = -\alpha_2/3 $ and the value for $\sigma_3$
given in Appendix \ref{App1}.

The first term of the expansion (\ref{expandQ0inX}),
$Q_0^{(1)}(\tau)=\cos(\tau)$, represents the linear (harmonic)
part of the in-phase motion. Higher order
corrections in $X^n$, $n > 1$ (i.e. $\varepsilon^{(n/2)}$) stem from the
nonlinearity of the onsite potential. It is important to notice that
they appear also for isochronous onsite potentials. In other words, 
{\em isochronous motions
are stricto sensu anharmonic}.

Reinstating the results for the $Q_0^{(n)}(\tau)$ in (\ref{expandQ0inX})
provides the general perturbative expression for the continuation of the linear
in-phase motion, that is, a perturbative expression for the {\em nonlinear
in-phase mode}.         

\subsection{Out-of-phase mode (orbit II)} \label{subsecoup}
\subsubsection{Equations of motion}

If the potential $V(x)$ is not symmetric  
($V(-x) \neq V(x)$), its Taylor expansion around 0 contains at
least one nonzero odd coefficient.
This has no influence on the previous result concerning the in-phase
motion because any oscillator of the chain performs the same motion in the same
time. This reduces the set of $N$ equations (\ref{eqmotQ}) 
to a single one (\ref{inp}), representing the 
equation of motion of a single oscillator in the
onsite potential $V$. But as soon as we are interested in an out-of-phase like
motion, we have to consider a dimerization of the chain, each dimer being made
of two neighbouring units oscillating in opposite phase. 
The lack of symmetry of $V$ 
induces two different motions to the right and to the left. 
This prevents us from finding a pure out-of-phase solution to
(\ref{eqmotQ}) which would imply $Q_q = Q_{\pi} \delta_{q,\pi}$ or in real
space $x_{2n}=-x_{2n+1}$. Instead, we can look for a solution of the type
\begin{equation}
Q_q = Q_0 \delta_{q,0} + Q_{\pi} \delta_{q,\pi} \label{OrbII}
\end{equation} 
involving both in- and out-of-phase variables, the 
others being zero. Using (\ref{xofQ}), we obtain 
$x_n = Q_0+(-1)^n Q_{\pi}$ or $x_{2n}=Q_0+Q_{\pi}$ and
$x_{2n+1}=Q_0-Q_{\pi}$. Adding and substracting the equations of motion for 
$x_{2n}$ and $x_{2n+1}$, we finally get 
\begin{eqnarray} 
&& \ddot{Q}_0 + \frac{1}{2}\left[ V'(Q_0+Q_{\pi}) + V'(Q_0-Q_{\pi})\right] =
0\, ,
\nonumber \\  
&& \ddot{Q}_{\pi} + \frac{1}{2}\left[ V'(Q_0+Q_{\pi}) - V'(Q_0-Q_{\pi})\right]
+ W'(2Q_{\pi}) - W'(-2Q_{\pi}) = 0 . \label{oop}
\end{eqnarray}

The total energy of the system evolving according to orbit II is
\begin{eqnarray}
E_{II} &=& H(\{ x_{2n}=Q_0+Q_{\pi},x_{2n+1}=Q_0-Q_{\pi}\}) \nonumber \\
&=& \frac{N}{2} \left(\dot{Q}_0^2+\dot{Q}_{\pi}^2
+V(Q_0+Q_{\pi})+V(Q_0-Q_{\pi})+W(2Q_{\pi})+ W(-2Q_{\pi}) \right) . 
\label{Eoop}
\end{eqnarray} 

\subsubsection{Solution at low energy} \label{solople}

System (\ref{oop}) represents a dimer. Applying the 
method of the first section, we are able to derive its time-periodic 
solution at low energy by requiring that the corresponding
orbit converges towards the linear out-of-phase mode as the energy vanishes. 
However, at variance with the in-phase mode, we cannot
provide an explicit expression for its period in terms of the
Taylor coefficients of $V$ and $W$. Removing the secular terms from
(\ref{oop}) yields simultaneously the expressions for the period and for the
motions $Q_0(\tau)$ and $Q_{\pi}(\tau)$. 

Contrary to the previous case, we won't use an energy expansion for
the diverse quantities to be calculated but merely a {\em small amplitude}
expansion. The small amplitude is denoted by $y$. We define a 
dimensionless time $\tau = \omega (y) t$ where $\omega (y)$ stands for
the frequency of orbit II as a function of its amplitude $y$. The
corresponding period is denoted by $T(y)$. Moreover, we define the two
dimensionless potentials 
\begin{equation} \label{VWtilde}
 \tilde{V}(x) = \frac{V(x)}{v_2+4\phi_2} = \sum_{n=2}^{\infty}
 \frac{\alpha_{n-1}}{n} x^n\ \ \ ;\ \ \ 
\tilde{W}(x) = \frac{W(x)}{v_2+4\phi_2} = \sum_{n=2}^{\infty}
 \frac{\beta_{n-1}}{n} \left(\frac{x}{2}\right)^n \, , 
\end{equation} 
as well as a dimensionless period $\tilde{T}(y) = \sqrt{v_2+4\phi_2}\,
T(y)/2\pi$. Notice the slightly different definition of the coefficients
$\alpha_n = v_{n+1}/(v_2+4\phi_2)$ of this section as compared to the previous
one. We nevertheless keep the same notation for the sake of clarity given that
these coefficients are still related to the onsite potential $V(x)$.
Functions $Q_0$, $Q_{\pi}$ and $\tilde{T}(y)$ are expanded as follows,
\begin{equation}
 Q_0(y,\tau) = \sum_{n=1}^{\infty} Q_0^{(n)}(\tau)\, y^n\ \ \ ;\ \ \ 
Q_{\pi}(y,\tau) = \sum_{n=1}^{\infty} Q_{\pi}^{(n)}(\tau)\, y^n \ \ \ ;\
\ \ \tilde{T}(y) = \sum_{n=1}^{\infty} \tilde{T}_n \, y^n  
\, . \label{expandQ0QpiTiny} 
\end{equation}  
Equations of motion (\ref{oop}) now read
\begin{eqnarray} 
&& \ddot{Q}_0 + \frac{\tilde{T}^2(y)}{2}\left[ \tilde{V}'(Q_0+Q_{\pi}) + \tilde{V}'(Q_0-Q_{\pi})\right] =
0\, ,
\nonumber \\  
&& \ddot{Q}_{\pi} + \tilde{T}^2(y)\left\{ \frac{1}{2}\left[
    \tilde{V}'(Q_0+Q_{\pi}) - \tilde{V}'(Q_0-Q_{\pi})\right] 
+ \tilde{W}'(2Q_{\pi}) - \tilde{W}'(-2Q_{\pi})\right\} = 0 . \label{oopad}
\end{eqnarray}     
The double dot denotes the derivative with respect to $\tau$. 
Inserting (\ref{expandQ0QpiTiny}) in the previous system and collecting
the terms of same order in $y$ gives rise to a set of differential
equations involving the functions $ Q_0^{(n)}(\tau)$, $
Q_{\pi}^{(n)}(\tau)$ as well as the unknowns $\tilde{T}_n$ (to be
determined by requiring the removal of secular terms). Solving them, it
is not difficult to obtain the following general features for the
motion: By choosing a proper origin of time the solution can be made
time-reversal symmetric (the only nonzero Fourier coefficients of 
$Q_0^{(n)}(\tau)$ and $Q_{\pi}^{(n)}(\tau)$ are even). Moreover,
$Q_0^{(n)}(\tau)$ and $Q_{\pi}^{(n)}(\tau)$ are respectively even and
odd in $y$. Finally, $\tilde{T}(y)$ is also even in $y$. Thus, we have
\begin{equation}
 Q_0(y,\tau) = \sum_{n=1}^{\infty} Q_0^{(2n)}(\tau)\, y^{2n}\ \ \ ;\ \ \ 
Q_{\pi}(y,\tau) = \sum_{n=0}^{\infty} Q_{\pi}^{(2n+1)}(\tau)\, y^{2n+1} \ \ \ ;\
\ \ \tilde{T}(y) = 1 + \sum_{n=1}^{\infty} \tilde{T}_{2n} \, y^{2n}  
\, . \label{expandQ0QpiTinygen} 
\end{equation}    
where
\begin{eqnarray} 
&& Q_{\pi}^{(1)}(\tau) = \cos(\tau), \nonumber \\ 
&& Q_0^{(2)}(\tau) =
\frac{1}{2}\alpha_2\left(\frac{\cos(2\tau)}{4-\alpha_1} - 
\frac{1}{\alpha_1}\right),
\nonumber \\  
&& Q_{\pi}^{(3)}(\tau) =
\frac{1}{32}\left[(\alpha_3+\beta_3)-\frac{2\alpha_2^2}{\alpha_1-4}\right]
\cos(3\tau), \nonumber \\
&& \tilde{T}_2 =
\frac{1}{4}\frac{(3\alpha_1-8)\alpha_2^2}{(\alpha_1-4)\alpha_1}
-\frac{3}{8}(\alpha_3+\beta_3). \label{solQ0QpiT}
\end{eqnarray}
to give a few. Notice that the even coefficients $\beta_{2n}$ are
absent from the equations of motion (\ref{oopad}) and consequently
from the expressions (\ref{solQ0QpiT}). $\beta_{1}$ has been eliminated thanks
to the relation $\beta_1+\alpha_1=1$, which follows from the definition of
these coefficients in terms of $v_2$ and $\phi_2$ (see eqs. (\ref{expandVW})
and (\ref{VWtilde})). 

Reinstating properly the expressions for $Q_0(y,\tau)$ and
$Q_{\pi}(y,\tau)$ in (\ref{Eoop}) allows us to derive a perturbative
expansion for the energy density $\varepsilon = E_{II}/N$ 
of the nonlinear out-of-phase mode in terms of its amplitude $y$.
As for the in-phase mode, it is convenient to define a quantity $Y =
(2\varepsilon/(v_2+4\phi_2))^{1/2}$ 
playing the same role as $X$ in the previous section. We then obtain
\begin{equation}
 \frac{2\varepsilon}{v_2+4\phi_2} \equiv Y^2 = y^2 +
 \left[\frac{9}{16}(\alpha_3+\beta_3)-\frac{1}{8}
\frac{(9\alpha_1^2-68\alpha_1+96)\alpha_2^2} 
{\alpha_1(\alpha_1-4)^2}\right]\, y^4 + {\cal O}\left(y^6\right)    
\, . \label{expandnrjiny} 
\end{equation}     
The first term of this expansion corresponds to the harmonic limit.

\section{Isochronism} \label{Sec3}

In the previous section, we have described the way to obtain 
a perturbative energy expansion
for both the in- and out-of-phase motions, as well as for their
respective periods. We are thus in a position to express the conditions
required for a (partial) isochronism of these modes. For an
isochronism of order $n$, this is easily achieved by cancelling all
coefficients of the energy expansion for the period up to order $n-1$.

\subsection{In-phase mode}

Due to its integral representation, the period of the in-phase mode can be
explicitly expanded as a power series in energy (see Appendix \ref{App1}). 
Obtaining an in-phase mode isochronous up to order $n$ 
then amounts to zeroing the coefficients $\sigma_{2k+1}$, $1 \leq k
\leq n-1$. This has been done in Appendix \ref{App1} up to order
$n=4$. The set of equations thus derived induces some constraints 
on the Taylor ($v_k$ or $\alpha_{k-1}$) coefficients of the potential $V$. It
leaves nevertheless an entire freedom on the choice of odd
coefficients ($v_{2k+1}$, $k \geq 1$) as already remarked in
\cite{Osypowski}. For a (1D) potential isochronous up to order $n$,
the even coefficients $v_{2k}$ are then determined by the odd ones
$v_{2k+1}$ up to $k=2n$. The rest of the expansion is free.

To illustrate this, let us derive the most general expansion of a
(${\cal C}^{\infty}$) 1D potential isochronous up to order 4. 
Using the results of appendix \ref{App1} and $V(x) = \omega^2 \tilde{V}(x)$, 
we obtain:
\begin{eqnarray}
\tilde{V}(x) & = & \frac{1}{2}\, x^2 + \frac{\alpha_2}{3}\, x^3  
+ \frac{5}{18}\alpha_2^2 \, x^4 
+  \frac{\alpha_4}{5}\, x^5 +\frac{1}{6}\left( -\frac{56}{27}\alpha_2^4 + 
\frac{14}{5}\alpha_4 \alpha_2\right)\, x^6 \nonumber \\
& + & \frac{\alpha_6}{7}\, x^7 + \frac{1}{8}\left( \frac{24}{7}\alpha_6 \alpha_2  
-\frac{592}{45}\alpha_4 \alpha_2^3 + \frac{36}{25}\alpha_4^2 
+ \frac{848}{81}\alpha_2^6 \right)\, x^8
+  \sum_{k=9}^{\infty} \frac{\alpha_{k-1}}{k}\, x^k
\end{eqnarray}
This expression is equivalent 
\footnote{Two misprints have to be
  corrected in this expression. Using the notations of
  \cite{Osypowski}, the last coefficient of order 6 reads $-7/2 \, b_1^4$ and
the last coefficient of order 8, $477/16 \, b_1^6$} to Eq. (34) of
  \cite{Osypowski}. 
  
\subsection{Out-of-phase mode} \label{isooprel} 

As for the out-of-phase motion, no explicit (integral) expression is
available for the period. However, we have shown in the previous
section how to calculate it as a function of the small amplitude $y$ by
removing the secular terms at each step of the calculation. The
period could of course be converted as a series in energy rather
than expressed as a series in amplitude. Such a transformation would be
pointless however. Indeed, even if an isochronism of order
$n$ has been previously defined through $T = T_0 + T_{2n}
\varepsilon^n + o(\varepsilon^n)$ as the 
energy tends to zero, the dependence of the amplitude on energy
(\ref{expandnrjiny}) shows 
that this statement is equivalent to $T = T_0 + T_{2n}
[(v_2+4\phi_2)/2]^n y^{2n} + o(y^{2n})$. It is then equivalent to zero 
the coefficients of the $y$ or $\varepsilon$ expansions up to order
$n-1$ as it produces the
same constraints on the Taylor coefficients of the potentials $V$ and
$W$, although these two series have different coefficients.

Let us finally provide the reader with an example to illustrate the
method described in the previous section. If we require the
out-of-phase mode to be isochronous up to order 3, we obtain the
following contraints on the coefficients $\alpha$ and $\beta$ (see
eqs. (\ref{VWtilde}) for their definition):
\begin{eqnarray}
\beta_3 &=& -\alpha_3 +
\frac{2}{3}\frac{(3\alpha_1-8)\alpha_2^2}{(\alpha_1-4)\alpha_1}
\nonumber \\
\beta_5 &=& -\alpha_5
+\frac{2}{15}\frac{(96+15\alpha_1^2-56\alpha_1)}
{\alpha_1^3(\alpha_1-4)^2}\alpha_2^4 
-\frac{9}{5}\frac{(32+5\alpha_1^2-24\alpha_1)}
{\alpha_1^3(\alpha_1-4)^2}\alpha_1\alpha_2^2\alpha_3
+\frac{6}{5}\frac{(5\alpha_1-12)}{\alpha_1^3(\alpha_1-4)}\alpha_1^2\alpha_2
\alpha_4 \label{betaiso} 
\end{eqnarray}  
Notice that for the sake of simplicity, we have kept the same notation
$\alpha_k$ for the coefficients of $\tilde{V}$ for both modes. However,
the definition of these coefficients is mode dependent as they
represent $v_{k+1}/v_2$ for the in-phase mode and
$v_{k+1}/(v_2+4\phi_2)$ for the out-of-phase mode. As already
displayed in the above relations, it can be shown in
general that the isochronism of orbit II is easily expressed through a
relation of the type $\beta_{2n+1} = f(\{\alpha_k\})$ ($k \in
\{1,\dots,2n+1\}$). This means that once the coefficients
$\{\alpha_k\}$ have been chosen for the onsite potential $V$, the
isochronism of the out-of-phase mode fixes the even part of the
interaction potential. The odd part of $W$ remains completely free as it does
not enter the equations of motion.

\section{Tangent bifurcation of orbit I and II} \label{Sec4}

\subsection{Statement of the problem}

The system of equations (\ref{eqmotQ}) is of the form $\ddot{Q}+F(Q)=0$, where
$Q$ and $F(Q)$ denote two vectors of components $Q_q$ and $F_q(Q)$
respectively, $q \in \{0,2\pi/N,\ldots,2(N-1)\pi/N \}$. A perturbation $\eta$
of the system around the solution $Q$ gives rise to the following variational system
\begin{equation} \label{vareqeta}
 \ddot{\eta}+DF(Q)\eta=0
\end{equation}
where $DF(Q)$ is the Jacobian matrix of $F$
evaluated in $Q$ whose components are $DF_{qk}(Q)=\frac{\partial F_q}{\partial
Q_k} (Q)$.     

To evaluate the Jacobian matrix, we use (\ref{Fvec}) and find
\begin{eqnarray}
\frac{\partial F_q}{\partial Q_k} (Q) = 
\frac{1}{N} \sum_{n=1}^{N} e^{i(q-k)n} &&  \left[    
V" \left( \sum_{q'} e^{-iq'n} Q_{q'} \right) +
(1-e^{ik}) W" \left( \sum_{q'} (1-e^{iq'}) e^{-iq'n} Q_{q'} \right) \right. 
\nonumber \\
 && \left.  -
(e^{-ik}-1) W" \left( \sum_{q'} (e^{-iq'}-1) e^{-iq'n} Q_{q'} \right) \right] .
\label{JacM}
\end{eqnarray}
We will now use the method already described in \cite{Fla96} in order
to obtain the energy thresholds, if any, at which orbits I and II
become unstable and bifurcate tangently to give rise to other types
of periodic orbits which break the translational invariance of the
lattice. Before to proceed, let us briefly recall how the method
works.

Once the periodic solution for $Q$ has been introduced in the Jacobian matrix
$DF(Q)$, the variational system (\ref{vareqeta}) presents itself as a vectorial
Hill's equation for the perturbation $\eta$. This type of systems is
known as {\em parametrically excited} as the Jacobian
matrix generally depends on several parameters. In our case, such
parameters are the energy (or the amplitude) of the solution
$Q$ as well as the frequencies of the modes we are interested in. 
Once expanded as a Fourier series, the Jacobian matrix
may be decomposed into a static (dc-) part (its zero mode) and a driving
(ac-) part. 

A paradigmatic example of Hill's equation is the Mathieu
equation $\ddot{x}(t)+(\delta +2\epsilon \cos(2t)) x(t)=0$. For a
comprehensive treatment of this equation the reader is invited to
consult \cite{Nayfey}, for example. In one
dimension (the Jacobian matrix is reduced to a single element in this
case), the parameter $\delta$ plays the role of the static part and
$2\epsilon \cos(2t)$ the role of the driving. It is known from the
stability analysis of this equation that the behaviour of its solution
varies according to the values of the parameters $\delta$ and
$\epsilon$. The solution can be either stable, unstable or
periodic. In the $\epsilon \delta$-plane ($\delta$ as $x$-axis and
$\epsilon$ as $y$-axis), the regions of instability  
present themselves as tongues (the so-called Arnold's tongues)
starting from the $\delta$-axis at the values $\delta_n = n^2$, $n \in \N$ and
widening as $\epsilon$ increases. In these regions, the motion is
unbounded whereas outside, it is stable (bounded). 

Of particular
interest are the boundaries of such regions called {\em transition
  curves} that separate stable from unstable motions. Along this
curves, the solution is periodic of period $\pi$ ($n$ even) or $2\pi$
($n$ odd). For small $\epsilon$ values, a perturbative treatment of the
Mathieu equation which consists in expanding both $x(t)$ and the
parameter $\delta$ as series in $\epsilon$ allows for the 
determination of the transition curves of the form $\delta_n = n^2 +
\sum_l A^{s/a}_{nl} \epsilon^l$. The coefficients $A^{s/a}_{nl}$
depend on the tongue ($n$) as well as on the {\em branch} (that is,
the boundary) we are interested in. It can be shown that one of these
branches is related to time reversal symmetric solutions $\eta$ (denoted
by the subscript $s$) whereas the second one is associated with time reversal
antisymmetric solutions ($a$).  
     
Let us suppose now that we fix the value of $\delta$ close to a
transition point $\delta_n=n^2$. At $\epsilon=0$, the point corresponding
to the state of the system in the parameter space is located in a
region of stability. Let us increase the value of $\epsilon$ at fixed
$\delta$. If the corresponding vertical line crosses the transition
curve nearby, the solution $x(t)$ becomes unstable above the crossing
point. And right at the intersecting point, the solution is periodic.
Although the variational equation (\ref{vareqeta}) is more complex 
than the Mathieu equation, we will proceed along
the same lines as those described above to derive the energies 
at which the modes (in- or-out-of-phase) become unstable and bifurcate.

\subsection{In-phase mode}

\subsubsection{Stability analysis.}
Evaluated along orbit I, the Jacobian matrix (\ref{JacM}) is
diagonal
\begin{equation} 
DF_{qk}(Q_I) = \left[ V"(Q_0) +4 \phi_2 \sin^2 \left(\frac{k}{2}\right)
\right] \delta_{q,k} \, .\label{JacI}
\end{equation} 
All perturbations decouple from each other and their equations of motion
are 
\begin{equation} 
\ddot{\eta}_q + \left[ V"(Q_0) +4 \phi_2 \sin^2 \left(\frac{q}{2}\right)
\right] \eta_q = 0 \, . \label{pertI} 
\end{equation} 
As already stated in \cite{Fla96},
the perturbation $\eta_0$ describes the continuation of orbit I along
itself. It cannot be responsible for a bifurcation of $Q_0$ as it
simply operates a shift in time or modifies the energy (or the
frequency) along the one-parameter family. We then look for the 
perturbation able to give rise to the required tangent
bifurcation. The first to occur will be for the closest 
(linear) frequency to the linear in-phase frequency,
that is for $q_c=2\pi/N$.

Let us now rewrite the system made of the dimensionless equations of
motion for $Q_0(\tau)$ and $\eta_{q_c}(\tau)$. We obtain
\begin{eqnarray} 
&& 
\frac{\partial^2 Q_0(X,\tau)}{\partial \tau^2} + \tilde{T}^2(X)
\tilde{V}'\left(Q_0(X,\tau)\right)=0, \label{dimlessinp2} \\
&& \frac{\partial^2 \eta_{q_c}(X,\tau)}{\partial \tau^2} +
\tilde{T}^2(X) \left[
\tilde{V}''\left(Q_0(X,\tau)\right)-\Delta \right] \eta_{q_c}(X,\tau)
=0, \label{dimlessinppert} 
\end{eqnarray} 
where 
\begin{equation} \label{Deltainp}
\Delta = - \frac{4\phi_2}{v_2} \sin^2\left(\frac{\pi}{N}\right) =
\frac{\omega^2_{0,0}-\omega^2_{q_c,0}}{v_2} .
\end{equation}
The notations used above are the same as in section
\ref{subsecinp}. The energy dependence of all quantities has been
emphasized through $X$. 

\subsubsection{Arnold tongue corresponding to a tangent bifurcation.}

Recall that due to the time scaling $\tau =
\omega (\varepsilon) t$ the period of $Q_0(X,\tau)$ is now $2\pi$. So
that looking for a tangent bifurcation of this orbit through
$\eta_{q_c}(X,\tau)$ requires $\eta_{q_c}(X,\tau+2\pi)=
\eta_{q_c}(X,\tau)$ as well. However, this condition in itself, although
necessary, is not sufficient for our purpose. We have indeed to
require the frequency of $\eta_{q_c}(X,\tau)$ to be the same as (and not
a multiple of) the driving frequency {\em in the limit of vanishing energy}
(or as $X \rightarrow 0$).  
In other words, expanding $\eta_{q_c}(X,\tau)$ as a Fourier series, we
must have
\begin{equation} \label{condetaFs}
\eta_{q_c}(X,\tau)= \sum_{n \in \Z} A_n(X) e^{i n \tau} \ \ \
\text{with}\ \ \  \forall |n| \neq 1 \, ,\ A_n(X)
\xrightarrow[X \rightarrow 0]{}  0 \, .
\end{equation}
This implies that we have to investigate the {\em first 
instability zone} of the Hill's equation (\ref{dimlessinppert}).

At a strictly zero energy ($X=0$), this imposes $\Delta=0$ for in this
limit, (\ref{dimlessinppert}) becomes 
$$ \ddot{\eta}_{q_c} + \left[1-\Delta \right] \eta_{q_c} =0 \, ,$$
(we again use the dot as a derivative with respect to $\tau$). The
previous equation  has a frequency equal to 1 for $\Delta=0$
only. Thus, in a diagram $(\Delta,\varepsilon)$, the Arnold tongue we
are concerned with starts from the point $(\Delta=0,\varepsilon=0)$. 

To evaluate the boundaries of this instability zone, we need to solve 
(\ref{dimlessinppert}) by expanding $\eta_{q_c}(X,\tau)$ and $\Delta$ as
power series in $X$:
\begin{equation} \label{etaDeltapsX}
\eta_{q_c}(X,\tau) = \sum_{m=0}^{\infty} \eta^{(m)}_{q_c}(\tau)\, X^m
\ \ \ ; \ \ \ \Delta = \sum_{m=0}^{\infty} \Delta^{(m)} \, X^m \, .
\end{equation}  
As for the Mathieu equation, 
removing the secular terms from (\ref{dimlessinppert}) 
gives rise to two critical curves $\Delta_s(X)$ and $\Delta_a(X)$
associated respectively with the periodic time reversal {\em symmetric}
solution $\eta_{q_c,s}(X,\tau)=\eta_{q_c,s}(X,-\tau)$ and  the
periodic time reversal {\em antisymmetric} solution
$\eta_{q_c,a}(X,\tau)=-\eta_{q_c,a}(X,-\tau)$.

\subsubsection{First transition curve.}

Differentiating (\ref{dimlessinp2}) with respect to $\tau$ and
dividing by $X$ shows that
\begin{equation} \label{sinbrancheq}
\frac{\partial^2 }{\partial \tau^2} \left(\frac{\dot{Q}_0}{X}\right)+ \tilde{T}^2(X)
\tilde{V}''\left(Q_0(X,\tau)\right)\left(\frac{\dot{Q}_0}{X}\right) =
0\, .
\end{equation}
As  $Q_0(X,\tau)$ is time reversal symmetric and equivalent to  $\cos(\tau) X
+ {\cal O}(X)$ when $X$ tends to zero (see \eqref{expandQ0inX} and
\eqref{Q0123inp}), $\dot{Q}_0(X,\tau)/X$ is time 
reversal antisymmetric and satifies (\ref{condetaFs}). We then
deduce that, for all $X$, $\dot{Q}_0(X,\tau)/X$ is an eigensolution of
(\ref{dimlessinppert}) for the eigenvalue $\Delta = 0$. In other
words, $\eta_{q_c,a}(X,\tau)=\dot{Q}_0(X,\tau)/X$ is the eigenfunction
corresponding to the eigenvalue $\Delta=0$ whatever the energy (or
$X$) is. We have then shown that the boundary $\Delta_a(X)$ of the
instability zone we are concerned with is the $\varepsilon$-axis of the
$\Delta - \varepsilon$ plane. 

\subsubsection{Second transition curve for a partial isochronism of order $n$.}

To calculate the second curve, $\Delta_s(X)$ related to the time reversal
symmetric solution $\eta_{q_c,s}(X,\tau)$ of (\ref{dimlessinppert}),
we differentiate (\ref{dimlessinp2}) with respect to $X$:
\begin{equation} \label{cosbrancheq}
\frac{\partial^2 }{\partial \tau^2} \left(\frac{\partial Q_0 }{\partial X}\right)+ \tilde{T}^2(X)
\tilde{V}''\left(Q_0\right) \left(\frac{\partial Q_0}{\partial
    X}\right) +\frac{d \, \tilde{T}^2(X)}{d X} \tilde{V}'\left(Q_0\right)=
0\, .
\end{equation}
This equation is not of the Hill type because of its last term which
is not linear in $\partial Q_0 /\partial X$. 

Suppose now that the potential $V$ is isochronous up to order $n$. By definition, we
have $\tilde{T}^2(X)=1+2\tilde{T}_{2n}\, X^{2n}+ o(X^{2n})$. That is
$\frac{d \tilde{T}^2}{d X} = 4n \tilde{T}_{2n} X^{2n-1}+ o(X^{2n-1})$
and $\tilde{V}'\left(Q_0\right)=X Q_0^{(1)}(\tau)+o(X^2)$. The last
term is then of order $X^{2n}$.

This means that, up to order $X^{2n-1}$, equation (\ref{cosbrancheq})
reads
\begin{equation} \label{cosbrancheqorderk}
\left[ \frac{\partial^2 }{\partial \tau^2} 
\left(\frac{\partial Q_0 }{\partial X}\right)+ \tilde{T}^2(X)
\tilde{V}''\left(Q_0\right) \left(\frac{\partial Q_0}{\partial
    X}\right) \right]_k = 
0\, ,\ \ 0 \leq k \leq 2n-1\, .
\end{equation}
(Here and hereafter, the symbol $[f(X)]_k$ will denote the terms of order
$X^k$ of $f(X)$, $[f(X)]_k= \frac{1}{k!} \frac{d^k f}{d X^k}
(0)$). Since $\partial Q_0/ 
\partial X = \cos(\tau) +{\cal O}(X)$ as $X \rightarrow 0$, it
satisfies (\ref{condetaFs}) and is moreover time reversal
symmetric. Now, because of  (\ref{cosbrancheqorderk}), up to order
$X^{2n-1}$, $\partial Q_0/\partial X$ represents the time reversal symmetric
eigenfunction $\eta_{q_c,s}(X,\tau)$ of (\ref{dimlessinppert}) for an
eigenvalue $\Delta_s$ which, up to this order, is equal to 0. This means
that
\begin{equation} \label{DeltasEtas}
\Delta_s(X) = \sum_{m=2n}^{\infty} \Delta_s^{(m)} \, X^m \ \ \
\text{and}\ \ \ \eta_{q_c,s}^{(k)}(\tau) \equiv \left[
  \eta_{q_c,s}(X,\tau)\right]_k =  
\left[ \frac{\partial Q_0}{\partial
    X} \right]_k\, ,\ \ 0 \leq k \leq 2n-1\, .
\end{equation}
Let us write now 
\begin{equation} \label{eta2n}
\eta_{q_c,s}^{(2n)}(\tau) = \left[ \frac{\partial Q_0}{\partial
    X} \right]_{2n}+\zeta(\tau) \, ,
\end{equation}
where $\zeta(\tau)$ is a function to be determined. Taking into
account (\ref{DeltasEtas}) and (\ref{cosbrancheqorderk}), we are able
to calculate the terms of order $X^{2n}$ of (\ref{dimlessinppert}) and
obtain
\begin{eqnarray} \label{pert2n}
&& \left[\frac{\partial^2 \eta_{q_c}(X,\tau)}{\partial \tau^2} +
\tilde{T}^2(X) \left\{
\tilde{V}''\left(Q_0(X,\tau)\right)-\Delta \right\} \eta_{q_c}(X,\tau)\right]_{2n}
=0 \nonumber \\
&\Leftrightarrow&
\ddot{\zeta}(\tau)+\zeta(\tau)-\Delta_s^{(2n)}\,Q_0^{(1)}(\tau)-4n
\tilde{T}_{2n}\,Q_0^{(1)}(\tau) = 0 \, .
\end{eqnarray}
As $Q_0^{(1)}(\tau)= \cos(\tau)$, removing the secular terms from this
last equation requires $\Delta_s^{(2n)} = - 4 n
\tilde{T}_{2n}$. Hence, the equation of the second branch of the
instability zone
\begin{equation} \label{Delta2n}
\Delta_s(X)= - 4 n \tilde{T}_{2n} X^{2n} + {\cal O}(X^{2n}) \, .
\end{equation}

\subsubsection{Bifurcation energy.} \label{Bifenipsec}

Coming back to the definitions,
$\Delta=(\omega^2_{0,0}-\omega^2_{q_c,0})/v_2 $,  
$X=(2\varepsilon/v_2)^{1/2}$, $\tilde{T}^2(X) = v_2/\omega_0^2$ and
$\omega_0^2 = \omega_{0,0}^2 + \gamma_{0,n}\, \varepsilon^n +
o(\varepsilon^n)$ ($\omega_{0,0}^2=v_2$), we can use (\ref{Delta2n}) to obtain the leading
order expression for the energy $\varepsilon_0^{\rm bif}$ at which a
nonlinear in-phase mode isochronous up to order $n$ bifurcates:
\begin{equation} \label{nrjbifinp}
 \varepsilon_0^{(n)}
 =\left(\frac{\omega^2_{0,0}-\omega^2_{q_c,0}}{2n\gamma_{0,n}}\right)^{1/n}=
\left(\frac{-2\phi_2 \sin^2\left(\frac{\pi}{N}\right)}{n\gamma_{0,n}}\right)^{1/n}
 \, .
\end{equation}  

\begin{center}
\begin{figure} 
\includegraphics[height=0.6\linewidth,width=0.7\linewidth]{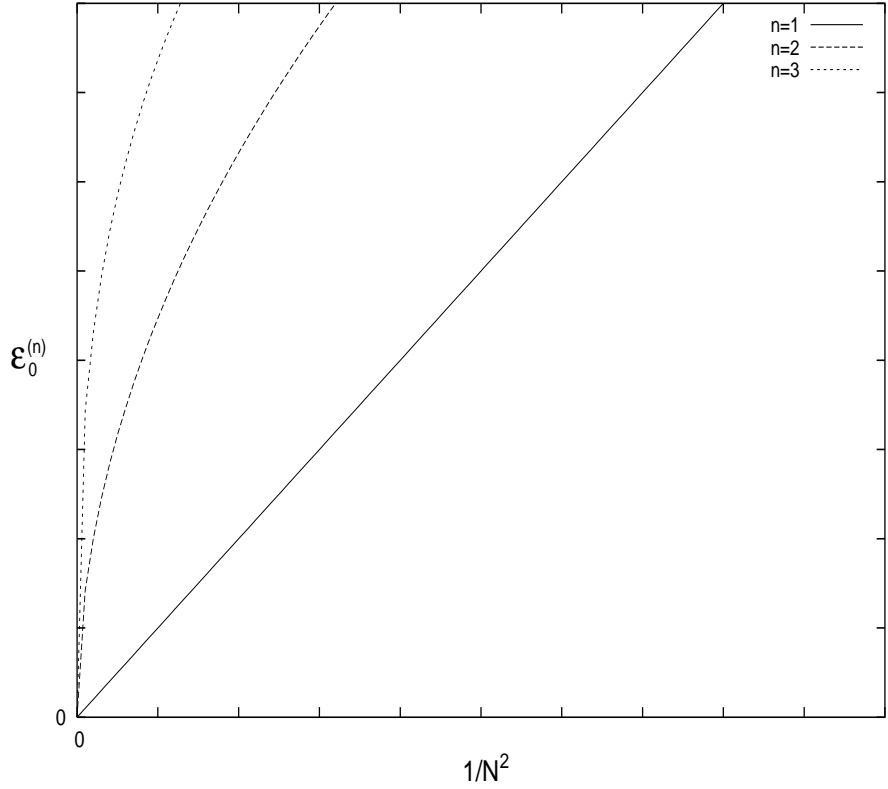}
\caption{Schematic representation of the bifurcation energy 
\eqref{dimlessinppert} as a function of the inverse squared 
number of oscillators $N$ for various degrees of isochronism of the in-phase
mode ($n=$1,2 and 3). For a given $n$, the 
instability zone lies between the energy-axis (first transition curve,
$1/N^2=0$) and the curve labelled by $n=i$, $i=$1,2 or 3. 
Below (above) the latter, the in-phase mode is stable (unstable). 
Right on a transition curve the solution is periodic. We see on this graph 
how the instability zone shrinks as $n$ increases. In the purely isochronous
case, the second transition curve merges with the first one on the
energy-axis and the corresponding instability zone disappears.} \label{fig1} 
\end{figure}
\end{center}

Let us first notice that the existence of a critical energy at which
the in-phase mode undergoes a bifurcation implies that $\gamma_{0,n}$ is
negative which, in turn, requires the corresponding frequency to decrease with
increasing energy. This condition, already known for non isochronous
potentials (or, in our terminology, isochronous up to order 1)
\cite{Fla96}, still holds in case of partial isochronism. 

As already mentioned in the introduction, expression \eqref{nrjbifinp} shows a
very deep relation between the low-energy behaviour of the frequency of the
in-phase mode and its bifurcation energy. We will see in the next section that
the same relation holds in fact for the out-of-phase mode. As $n$ may now take
on arbitrary positive integer values, \eqref{nrjbifinp} treats the case of
all possible analytic potentials with nonzero harmonic frequency ($v_2 >0$).
Its validity is however not restricted to such potentials and we remark that
it still holds for the class of non analytic potentials studied recently by
M. Kastner in relation with the possible 
existence of energy thresholds for discrete breathers
\cite{Kas04,Kaslong04}. From this point of view, this relation seems quite
general. 

\subsubsection{Examples.}

The coefficients $\gamma_{0,n}=-2v_2(2/v_2)^n\tilde{T}_{2n}$ are easily
obtained from the expression (\ref{per3}) for the period $T$. Indeed,
eq. (\ref{per3}) shows that $\tilde{T}_{2n}= \sigma_{2n+1}\,
(2n+1)!!/(2n)!! $. Then, zeroing the $\sigma_{2k+1}$, 
$k \in \{1,\dots,n-1\}$ as explained in Appendix \ref{App1} and
re-introducing the corresponding constraints in $\sigma_{2n+1}$,
allows us to derive the expression for $\tilde{T}_{2n}$ for a
potential isochronous up to order $n$.

\noindent
For $n=1$ (non isochronous potential), 
$$\tilde{T}_2 = \frac{3}{2}\, \sigma_3 =
\frac{3}{2}\left(\frac{5}{18}\alpha_2^2-\frac{1}{4}\alpha_3\right)
\Longrightarrow \gamma_{0,1} = \frac{3v_4}{2v_2}-\frac{5v_3^2}{3v_2^2} $$
then, 
\begin{equation}
\varepsilon^{(1)}_0 = \frac{12\phi_2
  v_2^2\sin^2(\frac{\pi}{N})}{10v_3^2-9v_2v_4}
\end{equation}
which is exactly the expression (3.13) of \cite{Fla96}.

\noindent
For $n=2$, 
$$\tilde{T}_2 = 0 \Longrightarrow \alpha_3=\frac{10}{9}\alpha_2^2 \Longrightarrow
\tilde{T}_4=\frac{15}{8}
\left(-\frac{1}{6}\alpha_5-\frac{28}{81}\alpha_2^4+\frac{7}{15}\alpha_4\alpha_2\right)  
\Longrightarrow \gamma_{0,2} =
-\frac{15}{v_2}\left(\frac{7}{15}\frac{v_3v_5}{v_2^2}-\frac{28}{81}\frac{v_3^4}{v_2^4}-\frac{1}{6}\frac{v_6}{v_2}\right)
\, .$$
Hence,
\begin{equation}
\varepsilon^{(2)}_0 =
\left(\frac{54v_2^3\phi_2}{378v_3v_5v_2^2-280v_3^4-135v_6v_2^3}\right)^{1/2}v_2
\sin \left(\frac{\pi}{N}\right) \, .
\end{equation}

\subsubsection{Pure isochronism} \label{sssIVB7}

Let us suppose now that the in-phase mode is purely
isochronous. Then, $\tilde{T}(X)=1$. Differentiating (\ref{dimlessinp2}) with
respect to $X$ as we have done in the previous section gives rise to 
 \begin{equation} \label{cosbrancheqpureiso}
\frac{\partial^2 }{\partial \tau^2} \left(\frac{\partial Q_0 }{\partial
X}\right)+  
\tilde{V}''\left(Q_0\right) \left(\frac{\partial Q_0}{\partial
    X}\right) = 0\, .
\end{equation} 
At variance with eq. (\ref{cosbrancheq}), this equation is of the Hill
type and $\frac{\partial Q_0}{\partial X}$ represents the time
reversal symmetric eigensolution of the instability zone under investigation. 
Its eigenvalue is $\Delta_s=0$ for all
$X$, i.e. whatever the energy is. We have thus found that, in case of
pure isochronism of the onsite potential, the boundaries $\Delta_a(X)$
and $\Delta_s(X)$ of the instability zone merge and make it
disappear. 
The merging of two transition curves is a phenomenon known in stability
theory as a {\em coexistence}. It occurs for instance in Ince's or Lam\'e's
equations and general conditions for its appearence are given in
\cite{Magnus79,Rand04}. We notice incidentally that for an "isotonic"
potential, $V(x)=\omega^2(x+1-1/(x+1))^2/8$, which is isochronous
(see for example \cite{Nieto81}), the
variational equation \eqref{dimlessinppert} reduces to an Ince's
equation. This allows for an explicit verification of the general result
stated above.
    
As a consequence of the disappearance of the instability region, 
{\em the in-phase mode never undergoes a tangent bifurcation}.
Remark that, at variance with the results obtained for a partial
isochronism, this result is {\em non perturbative}. 
Indeed, eq. \eqref{sinbrancheq}
and \eqref{cosbrancheqpureiso} are exact and their respective
eigensolutions $\eta_{q_c,a}(X,\tau)=\dot{Q}_0/X$ and 
$\eta_{q_c,s}(X,\tau)=\partial Q_0/\partial X$ as well.  
 
To conclude, let us notice that expression (\ref{nrjbifinp}) shows how the
instability zone shrinks as $n$ tends to
infinity (see also Fig. \ref{fig1}). 
Nevertheless, as it is valid in the limit of vanishing energies only
it couldn't have been used to prove the result above. 
       
\subsection{Out-of-phase mode}

\subsubsection{Stability analysis.}

Evaluated along orbit II, the Jacobian matrix (\ref{JacM}) now reads
\begin{eqnarray} 
DF_{qk}(Q_{II}) &=& 
\frac{1}{2} \Big[ V"(Q_0+Q_{\pi})+V"(Q_0-Q_{\pi})+4\sin^2 \frac{k}{2}\, \{ W"(2
 Q_{\pi})+W"(-2 Q_{\pi})\}\Big] \delta_{k,q} \nonumber \\
&+&
\frac{1}{2} \Big[V"(Q_0+Q_{\pi})-V"(Q_0-Q_{\pi})-2i\sin k \,\{ W"(2
 Q_{\pi})-W"(-2 Q_{\pi})\}\Big] \delta_{k,q+\pi}\, ,  \label{JacII}
\end{eqnarray} 
and the corresponding dynamics,
\begin{eqnarray} 
\ddot{\eta}_q &+& 
\frac{1}{2} \Big[ V"(Q_0+Q_{\pi})+V"(Q_0-Q_{\pi})+4\sin^2 \frac{q}{2}\, \{
W"(2 Q_{\pi})+W"(-2 Q_{\pi})\}\Big] \eta_q \nonumber \\
&+&
\frac{1}{2} \Big[V"(Q_0+Q_{\pi})-V"(Q_0-Q_{\pi})+2i\sin q \,\{ W"(2
 Q_{\pi})-W"(-2 Q_{\pi})\}\Big] \eta_{q+\pi} = 0 \, . \label{pertII}
\end{eqnarray} 

To evaluate the possible bifurcating energies in this case, we will
proceed along the same lines as for the in-phase mode. Notice, however,
that eqs. (\ref{pertII}) do not decouple in the present case and that,
due to the very last term, they are complex (real and imaginary parts
are not decoupled). To deal with
this first difficulty, we will write the system (\ref{pertII}) in terms of
real and imaginary parts. 

Similar to the in-phase mode, the out-of-phase mode will eventually undergo a first
tangent bifurcation via the perturbation $\eta_{q_c}$ whose frequency 
is the closest to $\omega_{\pi,0}$, that is, for $q_c=\pi-2\pi/N$. As
$N$ tends to infinity, $q=2\pi/N$ plays the role of the small parameter
in the variational equations. But at variance with
the in-phase variational equation where $\Delta \propto \sin^2(\pi/N)$ 
was the unique small parameter, eqs. (\ref{pertII}) possess two small
parameters through $\sin^2((q_c+\pi)/2) \sim (\pi/N)^2$ and
$\sin(q_c) \sim \pi/N$. Notice that these two parameters are not of the same order. 

In what follows, for the sake of clarity, we simplify further the
notations by using $\eta_{q_c}=r_{\pi}+ij_{\pi}$ and
$\eta_{q_c+\pi}=r_{0}+ij_{0}$ (with $i=\sqrt{-1}$)
rather than the lengthy values for $q_c$ and $q_c+\pi$. We work
with dimensionless equations and with the same notations as in section
\ref{subsecoup}. We take $\delta = 4 \sin^2(\pi/N)$ as small parameter.
 The quantities
$r_0$, $j_0$, $r_{\pi}$, $j_{\pi}$, $Q_0$, $Q_{\pi}$ and $\tilde{T}$
depend on the amplitude $y$ defined in \ref{subsecoup}. As the previous
quantities, the small parameter $\delta$ has to be expanded as a
power series in $y$. Moreover, we define 
\begin{eqnarray} \label{TermsT}
{\cal T}_1(\delta,\tau) &=&
\frac{\tilde{T}^2}{2}\left[\tilde{V}"(Q_0+Q_{\pi})+\tilde{V}"(Q_0-Q_{\pi})+(4-\delta)\, \{ \tilde{W}"(2 Q_{\pi})+\tilde{W}"(-2 Q_{\pi})\}\right]\, , \nonumber \\
{\cal T}_2(\tau) &=&
\frac{\tilde{T}^2}{2}\left[\tilde{V}"(Q_0+Q_{\pi})-\tilde{V}"(Q_0-Q_{\pi})\right]\, ,
\nonumber \\
{\cal T}_3(\delta,\tau) &=&
\tilde{T}^2\sqrt{\delta(1-\delta/4)} \left[\tilde{W}"(2 Q_{\pi})-\tilde{W}"(-2 Q_{\pi})\right]\, , \nonumber \\
{\cal T}_4(\delta,\tau) &=&  \frac{\tilde{T}^2}{2}\left[\tilde{V}"(Q_0+Q_{\pi})+\tilde{V}"(Q_0-Q_{\pi})+4\delta\, \{ \tilde{W}"(2 Q_{\pi})+\tilde{W}"(-2 Q_{\pi})\}\right] \, .
\end{eqnarray}  
System (\ref{pertII}) now reads (from now on, dots denote $\tau$ derivative),
\begin{equation} \label{Eqrj}
\left\{
\begin{array}{ccc}
\displaystyle{\ddot{r}_{\pi}+{\cal T}_1(\delta,\tau)r_{\pi}+{\cal
    T}_2(\tau)r_{0}-{\cal T}_3(\delta,\tau)j_{0}} &
= & 0,\\ 
& & \\
\displaystyle{\ddot{j}_{\pi}+{\cal T}_1(\delta,\tau)j_{\pi}+{\cal
    T}_2(\tau)j_{0}+{\cal T}_3(\delta,\tau)r_{0}} & 
= & 0,\\ 
& & \\
\displaystyle{\ddot{r}_{0}+{\cal T}_4(\delta,\tau)r_{0}+{\cal
    T}_2(\tau)r_{\pi}+{\cal T}_3(\delta,\tau)j_{\pi}} & 
= & 0,\\ 
& & \\
\displaystyle{\ddot{j}_{0}+{\cal T}_4(\delta,\tau)j_{0}+{\cal
    T}_2(\tau)j_{\pi}-{\cal T}_3(\delta,\tau)r_{\pi}} & = & 0 . 
\end{array}
\right.
\end{equation}
As the energy (amplitude $y$) tends to zero, $Q_{\pi}$ and $Q_{0}$
tend to zero as well and from \eqref{TermsT}, \eqref{expandQ0QpiTinygen} 
and \eqref{solQ0QpiT}, we obtain ${\cal
  T}_1(\delta,\tau) \rightarrow 1-\delta\beta_1/4$, ${\cal
  T}_4(\delta,\tau) \rightarrow \alpha_1+\delta\beta_1/4$ and ${\cal
  T}_2(\delta,\tau)$ and ${\cal T}_3(\delta,\tau)$ both tend to 0. 
As $\delta$ is a
small parameter, the only way to obtain a perturbation whose frequency
is exactly 1 (as for the mode itself) is to require $\delta=0$ for it gives
rise to ${\cal T}_1 =1$ which is precisely the desired frequency for
the perturbation $(r_{\pi},j_{\pi})$. Regarding $(r_{0},j_{0})$, as
${\cal T}_4 =\alpha_1 < 1$ in this limit, we have to require
$(r_{0}=0,j_{0}=0)$ in order zero in $y$. The origin of the
instability zone is thus located at $\delta=0,\varepsilon=0$ in the parameter
space.

\subsubsection{First transition curve.}
Now, proceeding as in the previous section, let us derive the
equations obeyed by ${\cal Q}_0 = \dot{Q}_0/y$ and ${\cal Q}_{\pi} =
\dot{Q}_{\pi}/y$.
\begin{equation} \label{sinbranchop}
\left\{
\begin{array}{ccc}
\displaystyle{\ddot{{\cal Q}}_{\pi}+{\cal T}_1(0,\tau){\cal Q}_{\pi} +
  {\cal T}_2(\tau){\cal Q}_{0}} & = & 0,\\ 
& & \\
\displaystyle{\ddot{{\cal Q}}_{0}+{\cal T}_4(0,\tau){\cal Q}_{0} +
 {\cal T}_2(\tau){\cal Q}_{\pi}} & = & 0. 
\end{array}
\right.
\end{equation}
As the out-of-phase mode has been choosen time reversal symmetric,
${\cal Q}_0$  and ${\cal Q}_{\pi}$ are time reversal
antisymmetric. Moreover, due to the asymptotic relations given above, 
they represent a motion of frequency 1 as $y \rightarrow 0$. Having 
the required properties, they form the antisymmetric eigensolution of the
Hill's 
system \eqref{Eqrj} for a value of $\delta$ equal to zero for all $y$
(or energy). Therefore, $(r_{\pi},j_{\pi}) \propto {\cal Q}_{\pi}$ and
$(r_{0},j_{0}) \propto {\cal Q}_{0}$. We thus deduce that, in a diagram
$(\delta,\varepsilon)$, one of the boundaries of the
Arnold tongue is the $\varepsilon$-axis itself which means that, similar 
to the result obtained for the in-phase mode, no
bifurcation can be expected from this branch.         

\subsubsection{Second transition curve for an isochronism of order $n$.}
\label{secIVC3} 

The second branch of the instability zone is
the curve associated with the time reversal symmetric
eigenfunction. Differentiating (\ref{oopad}) with respect to $y$, we
obtain
\begin{equation} \label{cosbranchop}
\left\{
\begin{array}{ccc}
\displaystyle{\frac{\partial^2}{\partial
  \tau^2}\left(\frac{\partial Q_{\pi}}{\partial y}\right) 
+{\cal T}_1(0,\tau)\left(\frac{\partial Q_{\pi}}{\partial y}\right)+
  {\cal T}_2(\tau)\left(\frac{\partial Q_{0}}{\partial
  y}\right)-\frac{d \ln \tilde{T}^2}{d y} \frac{\partial^2 Q_{\pi}}{\partial
  \tau^2}}& = & 0,\\ 
& & \\ 
\displaystyle{\frac{\partial^2}{\partial
  \tau^2}\left(\frac{\partial Q_{0}}{\partial y}\right) 
+{\cal T}_4(0,\tau)\left(\frac{\partial Q_{0}}{\partial y}\right)+
  {\cal T}_2(\tau)\left(\frac{\partial Q_{\pi}}{\partial
  y}\right)-\frac{d \ln \tilde{T}^2}{d y} \frac{\partial^2 Q_{0}}{\partial
  \tau^2}}& = & 0 \, .
\end{array}
\right.
\end{equation}
Again, the functions $\partial Q_{\pi}/\partial y$ and $\partial
Q_{0}/\partial y$ have the right time symmetry and Fourier
properties. Nevertheless, the system above is not a Hill's system due
to the last terms. For an out-of-phase mode isochronous up to order $n$,
$\tilde{T}(y)=1+\tilde{T}_{2n} y^{2n}+o(y^{2n})$, and we can evaluate the
order of the last terms in \eqref{cosbranchop}. Using
\eqref{expandQ0QpiTinygen} and \eqref{solQ0QpiT}, we find
\begin{equation} \label{leadordlastterms}
-\frac{d \ln \tilde{T}^2}{d y} \frac{\partial^2 Q_{\pi}}{\partial
  \tau^2}  =  4n\tilde{T}_{2n}\cos \tau y^{2n} +o(y^{2n})\ \ ;\ \ 
-\frac{d \ln \tilde{T}^2}{d y} \frac{\partial^2 Q_{0}}{\partial
  \tau^2}  = {\cal O}(y^{2n+1}) \, .
\end{equation}
Given this last result, we will first assume, and then verify a posteriori,
that the order of $\delta$ in $y$ is $2n$, hence
\begin{equation} \label{orderdelta}
\delta =  \delta^{(2n)} y^{2n} +o(y^{2n})\, .
\end{equation} 
We then obtain immediately
\begin{eqnarray} \label{asymptT134}
{\cal T}_1(\delta,\tau) &=& {\cal
T}_1(0,\tau)-\frac{\beta_1}{4}\delta^{(2n)}y^{2n} +o(y^{2n})\, , \nonumber \\
{\cal T}_2(\tau) &=& 2\alpha_2 \cos \tau y +o(y)\, , \nonumber \\
{\cal T}_3(\delta,\tau) &=& \sqrt{\delta^{(2n)}}\beta_2 \cos \tau 
y^{n+1} +o(y^{n+1}) \, ,\nonumber \\
{\cal T}_4(\delta,\tau) &=& {\cal
T}_4(0,\tau)+\frac{\beta_1}{4}\delta^{(2n)}y^{2n} +o(y^{2n}) \, .
. 
\end{eqnarray}
We deduce from these relations that system \eqref{Eqrj} verifies
\begin{equation} \label{Eqrjupon}
\left\{
\begin{array}{ccc}
\displaystyle{\ddot{r}_{\pi}+{\cal T}_1(0,\tau)r_{\pi}+{\cal
    T}_2(\tau)r_{0}} &
= & \displaystyle{o(y^{n})},\\ 
& & \\
\displaystyle{\ddot{j}_{\pi}+{\cal T}_1(0,\tau)j_{\pi}+{\cal
    T}_2(\tau)j_{0}} & 
= & \displaystyle{o(y^{n})},\\ 
& & \\
\displaystyle{\ddot{r}_{0}+{\cal T}_4(0,\tau)r_{0}+{\cal
    T}_2(\tau)r_{\pi}} & 
= & \displaystyle{o(y^{n})},\\ 
& & \\
\displaystyle{\ddot{j}_{0}+{\cal T}_4(0,\tau)j_{0}+{\cal
    T}_2(\tau)j_{\pi}} & = & \displaystyle{o(y^{n})} . 
\end{array}
\right.
\end{equation}
We remark now that the solution $(\eta_{q_c}(\tau),\eta_{q_c+\pi}(\tau))$ of
\eqref{pertII} is defined up to a global phase only (i.e. if
$(\eta_{q_c}(\tau),\eta_{q_c+\pi}(\tau))$ is a solution of \eqref{pertII},
$(e^{i\phi} \eta_{q_c}(\tau),e^{i\phi} \eta_{q_c+\pi}(\tau))$, $\phi \in \R$
is also a solution). Using this freedom, we solve \eqref{Eqrjupon} by
requiring 
\begin{equation} \label{rjordern}
r_{\pi}^{(k)} = \left[\frac{\partial Q_{\pi}}{\partial y}\right]_k \ \ ;\ \
r_{0}^{(k)} = \left[\frac{\partial Q_{0}}{\partial y}\right]_k \ \ ;\ \
j_{\pi}^{(k)} = 0 \ \ ;\ \
j_{0}^{(k)} = 0\ \ ,\ 1 \leq k \leq n\, ,
\end{equation}   
without loss of generality (the brackets have the same meaning as in the
previous section). Reinstating in \eqref{Eqrj} and using \eqref{asymptT134}
we find
\begin{equation} \label{rjorder2n}
r_{\pi}^{(k)} = \left[\frac{\partial Q_{\pi}}{\partial y}\right]_k \ \ ;\ \
r_{0}^{(k)} = \left[\frac{\partial Q_{0}}{\partial y}\right]_k \ \ ,\ 1 \leq k
\leq 2n-1\, . 
\end{equation}
As in the last section, we now solve the first and the third equation of
\eqref{Eqrj} at order $2n$ by introducing two functions $\zeta_{\pi}(\tau)$ and
$\zeta_{0}(\tau)$ according to
\begin{equation} \label{rorder2n}
r_{\pi}^{(2n)} = \left[\frac{\partial Q_{\pi}}{\partial y}\right]_{2n}
+\zeta_{\pi}\ \ ;\ \ 
r_{0}^{(2n)} = \left[\frac{\partial Q_{0}}{\partial y}\right]_{2n} +
\zeta_0 = \zeta_0\, . 
\end{equation}
>From \eqref{cosbranchop} and \eqref{leadordlastterms} we have 
\begin{equation} \label{cosbranchop2n}
\left\{
\begin{array}{ccc}
\displaystyle{\left[\frac{\partial^2}{\partial
  \tau^2}\left(\frac{\partial Q_{\pi}}{\partial y}\right) 
+{\cal T}_1(0,\tau)\left(\frac{\partial Q_{\pi}}{\partial y}\right)+
  {\cal T}_2(\tau)\left(\frac{\partial Q_{0}}{\partial
  y}\right)\right]_{2n} + 4n \tilde{T}_{2n} \cos \tau} & = & 0,\\ 
& & \\ 
\displaystyle{\left[\frac{\partial^2}{\partial
  \tau^2}\left(\frac{\partial Q_{0}}{\partial y}\right) 
+{\cal T}_4(0,\tau)\left(\frac{\partial Q_{0}}{\partial y}\right)+
  {\cal T}_2(\tau)\left(\frac{\partial Q_{\pi}}{\partial
  y}\right)\right]_{2n}}& = & 0 \, .
\end{array}
\right.
\end{equation}
and from \eqref{Eqrj} and \eqref{asymptT134}
\begin{equation} \label{rpir02n}
\left\{
\begin{array}{ccc}
\displaystyle{\left[\ddot{r}_{\pi} 
+{\cal T}_1(0,\tau)r_{\pi} +
  {\cal T}_2(\tau)r_{0}\right]_{2n} -\frac{\beta_1}{4} \delta^{(2n)} \cos
\tau} & = & 0,\\  
& & \\ 
\displaystyle{\left[\ddot{r}_{0} 
+{\cal T}_4(0,\tau)r_{0} +
  {\cal T}_2(\tau)r_{\pi}\right]_{2n}} & = & 0\, .
\end{array}
\right.
\end{equation}
Therefore,
\begin{equation} \label{zetapi0}
\left\{
\begin{array}{ccc}
\ddot{\zeta}_{\pi} 
+\zeta_{\pi} -\left(\frac{\beta_1}{4} \delta^{(2n)}+4n
\tilde{T}_{2n}\right)\cos \tau & = & 0,\\  
\ddot{\zeta}_{0} +\alpha_1 \zeta_{0} & = & 0\, .
\end{array}
\right.
\end{equation}
Removing the secular term from the first equation leads eventually to 
the equation of the second transition curve
\begin{equation} \label{relatdeltay}
\delta = -\frac{16n\tilde{T}_{2n}}{\beta_1} y^{2n} + o(y^{2n})\, .
\end{equation}
Discarding the solution of the homogenous equation \cite{Nayfey} 
gives $\zeta_{\pi}=0$. The second equation is satisfied by setting
$\zeta_{0}=0$ as well (because its frequency $\alpha_1 \neq 1$). 
Hence, $r_{\pi}^{(2n)} = \left[\frac{\partial Q_{\pi}}{\partial y}\right]_{2n}
$ and $r_{0}^{(2n)} = \left[\frac{\partial Q_{0}}{\partial y}\right]_{2n}$. 
Remark that these
results are correct provided we are able to justify a posteriori the initial
assumption \eqref{orderdelta} regarding the small parameter $\delta$. This
amounts to proving that the remaining equations for $j_{\pi}$ and $j_0$ in
\eqref{Eqrj} never develop any secular terms up to order $2n$. 
This is done in appendix \ref{App2} by showing that these equations 
verify the adequate solvability conditions. 
 
\subsubsection{Bifurcation energy.}

Using the leading order relation between
the amplitude and the energy density, $y^2 =
2\varepsilon/\omega_{\pi,0}^2+o(\varepsilon^2)$,
$\omega_{\pi,0}^2=v_2+4\phi_2$, together with the relation 
\begin{equation} \label{omegaop}
 \omega_{\pi}^2 = \omega_{\pi,0}^2 + \gamma_{\pi,n}\varepsilon^n +
 o(\varepsilon^n) \, , 
\end{equation} 
and noting that $\delta=4\sin^2(\pi/N)$,
$\tilde{T}_{2n}y^{2n}=-\gamma_{\pi,n}\varepsilon^n /2\omega^2_{\pi,0}
+o(\varepsilon^n)$, we finally obtain from \eqref{relatdeltay} 
\begin{equation} \label{nrjbifinop}
 \varepsilon_{\pi}^{(n)} = 
\left(\frac{\omega^2_{\pi,0}-\omega^2_{q_c,0}}{2n\gamma_{\pi,n}}\right)^{1/n}
= \left(\frac{2\phi_2
\sin^2\left(\frac{\pi}{N}\right)}{n\gamma_{\pi,n}}\right)^{1/n} \, , 
\end{equation}  
which is similar to the expression obtained for the in-phase mode. 
We notice this time that the existence of a critical energy at which
the out-of-phase mode undergoes a bifurcation implies that $\gamma_{0,n}$ is
positive. This requires the corresponding frequency to increase with
the energy. This condition, already known for non isochronous 
motions \cite{Fla96}, still holds in case of partial isochronism.

As seen in section \ref{solople}, the period of the motion is more
conveniently expressed as a series in the square of the amplitude, $y^2$, 
rather than as a series in the energy $\varepsilon$. For this reason, we also
provide the bifurcation energy in terms of the coefficients $\tilde{T}_{2n}$
defined in \eqref{expandQ0QpiTinygen},
\begin{equation} \label{nrjbifinopT}
 \varepsilon_{\pi}^{(n)} = 
 \frac{\omega^2_{\pi,0}}{2} \left(\frac{-\phi_2
\sin^2\left(\frac{\pi}{N}\right)}
{n\omega^2_{\pi,0}\tilde{T}_{2n}}\right)^{1/n} \, .    
\end{equation}  
 
\subsubsection{Examples.}
We are now in a position to give an explicit expression of
$\varepsilon_{\pi}^{(n)}$ for an out-of-phase mode isochronous up to order
$n$ in terms of the parameters $v_i$ and $\phi_i$ defining $V(x)$ and $W(x)$
(see \eqref{expandVW} and \eqref{VWtilde}). 
This amounts simply to finding the leading order of the
low-energy (or low-amplitude) 
behaviour of the corresponding frequency as explained in section
\ref{solople} (see also section \ref{isooprel}).   

For a non isochronous out-of-phase mode ($n=1$), from \eqref{solQ0QpiT} and
\eqref{nrjbifinopT}, we find
\begin{equation} \label{ebifopn1}
\varepsilon_{\pi}^{(1)} = \frac{4
(v_2+4\phi_2)\phi_2 }{3(v_4+16\phi_4)+
\frac{2v_3^2}{3v_2+16\phi_2}-\frac{4v_3^2}{v_2}}\, \sin^2
\left(\frac{\pi}{N}\right) 
\end{equation}
\noindent
Let us notice first that, when $v_3=0$ this expression reduces to
formula (3.20) of \cite{Fla96} obtained 
in the special case of a symmetric onsite potential $V(x)$. The 
correction introduced by the asymmetry of $V(x)$ (i.e. 
the term proportional to $v^2_3$ in the denominator of \eqref{ebifopn1}) has
the interesting feature to be always negative. Therefore, the following 
inequality 
\begin{equation} \label{ineqbifopn1}
v_3^2 < \frac{3v_2(v_4+16\phi_4)(3v_2+16\phi_2)}{2(5v_2+32\phi_2)}
\end{equation}
\noindent
has to be satisfied for the out-of-phase mode to undergo a tangent 
bifurcation. This corresponds, as we have seen in the section above, to
requiring that the frequency increases with the energy.

Another interesting result easily drawn from \eqref{ebifopn1} concerns the
case of partially isochronous {\em onsite} potentials. It is found in this
case that a certain amount of nonlinearity ($\phi_4$) in
the interaction potential is needed in order to ensure a bifurcation of the 
{\em out-of-phase mode}. Indeed, the relation between the
first coefficients of the Taylor expansion of $V(x)$ is $10 v_3^2 = 9 v_2
v_4$ in this case. The denominator of \eqref{ebifopn1} is then positive
provided 
\begin{equation} 
\phi_4 > \frac{1}{5}\frac{v_4 \phi_2}{3v_2+16\phi_2} \label{crline}
\end{equation}  
So that, in a chain of harmonically coupled partially
isochronous oscillators, no discrete breather (if any) stems
from the tangent bifurcation of the out-of-phase mode.
At the same time we can conclude, that breathers appear for
fully isochronous harmonic oscillators ($v_4=0$) when coupled
anharmonically ($\phi_4 > 0$). Indeed, it has been recently proved
that breathers exist and can be continued from zero anharmonic
coupling for harmonic oscillators \cite{FlachDorignac03}.

For a first degree of isochronism ($n=2$), we obtain from \eqref{nrjbifinopT}
\begin{equation} \label{nrjbifinopTn2}
 \varepsilon_{\pi}^{(2)} = 
 \frac{v_2+4\phi_2}{2} \left(\frac{-\phi_2}
{2(v_2+4\phi_2)\tilde{T}_{4}}\right)^{1/2}
\, \sin \left(\frac{\pi}{N}\right) \, ,    
\end{equation}  
where
\begin{eqnarray} \label{T4isoop}
\tilde{T}_4 =
\frac{(15\alpha_1^2+96-56\alpha_1)\alpha_2^4}{24\alpha_1^3(\alpha_1-4)^2}-
\frac{9(32+5\alpha_1^2-24\alpha_1)\alpha_3\alpha_2^2}
{16\alpha_1^2(\alpha_1-4)^2}  
+ 
\frac{3(5\alpha_1-12)\alpha_4\alpha_2}{8\alpha_1(\alpha_1-4)}- 
\frac{5}{16}(\alpha_5+\beta_5)
\end{eqnarray}
with $\alpha_n=v_{n+1}/(v_2+4\phi_2)$ and
$\beta_n=2^{n+1}\phi_{n+1}/(v_2+4\phi_2)$. Notice that, to enforce a partial
isochronism of the out-of-phase mode up to order $n=2$, 
we have used the first of the relations \eqref{betaiso} 
($\beta_3 =f(\{\alpha_j\})$) to derive \eqref{T4isoop}.
Again, a tangent bifurcation of this mode will
take place provided the inequality $\tilde{T}_4<0$ is fulfilled.   
 
\subsubsection{Pure isochronism} \label{sssIVC6}
 
Similar to the pure isochronism of the in-phase mode, the pure isochronism of
the out-of-phase mode leads to a merging of the two transition curves which
makes the instability zone disappear. This is clear from the fact
that, as $\tilde{T}(y)=1$, $\partial Q_{\pi}/\partial y$ and $\partial
Q_{0}/\partial y$ obey
\begin{equation} \label{cosbranchoppi}
\left\{
\begin{array}{ccc}
\displaystyle{\frac{\partial^2}{\partial
  \tau^2}\left(\frac{\partial Q_{\pi}}{\partial y}\right) 
+{\cal T}_1(0,\tau)\left(\frac{\partial Q_{\pi}}{\partial y}\right)+
  {\cal T}_2(\tau)\left(\frac{\partial Q_{0}}{\partial
  y}\right)} & = & 0,\\ 
& & \\ 
\displaystyle{\frac{\partial^2}{\partial
  \tau^2}\left(\frac{\partial Q_{0}}{\partial y}\right) 
+{\cal T}_4(0,\tau)\left(\frac{\partial Q_{0}}{\partial y}\right)+
  {\cal T}_2(\tau)\left(\frac{\partial Q_{\pi}}{\partial
  y}\right)} & = & 0
\end{array}
\right.
\end{equation}
and are thus exact solutions of the variational equations
\eqref{Eqrj} for $q=0$ (i.e. $(r_{\pi},j_{\pi}) \propto \partial
Q_{\pi}/\partial y$ and $(r_{0},j_{0}) \propto \partial
Q_{0}/\partial y$).
As a consequence, {\em if the out-of-phase mode is isochronous it doesn't
undergo a tangent bifurcation}.    

\section{Summary and Discussion} \label{Sec5}

We have examined so far the close link existing
between the possible tangent bifurcation of 
band edge modes and the
low-energy behaviour of their frequency. We have introduced the concept of
partial isochronism of order $n$ for these modes through the relation 
$\omega^2_q=\omega^2_{q,0}+\gamma_{q,n} \varepsilon^n +o(\varepsilon^n)$ ($q=0$
or $\pi$). By performing a linear stability analysis in the limit 
of small oscillations, we have derived a simple and general  
expression for the leading order of their bifurcation energy in terms 
of $\gamma_{q,n}$. We have shown that the calculation of 
$\gamma_{0,n}$ (in-phase mode) simply requires to invert a series and thus
reduces to a pure algebraic problem. The inversion of this series is 
easily implemented with the help of any software able to perform formal
calculations. The coefficient $\gamma_{\pi,n}$ related to the out-of-phase
mode may be derived by means of a Lindstedt-Poincar\'e expansion for the
motion. 

In addition, we have proved by means of an {\em exact} linear
stability analysis that, fully isochronous band edge modes 
(i.e. with an energy-independent frequency) never undergo a tangent 
bifurcation. At variance with the results quoted above, this one is 
{\em non perturbative}. 

In order to discuss the implications of these results on discrete breathers, we
shall now assume that, at least some of them stem from the tangent
bifurcation of the band edge modes investigated so far. To our knowledge, for
a Klein-Gordon lattice with smooth (say ${\cal C}^{\infty}$) 
linearizable onsite {\em and} 
interaction potentials ($\omega_{0,0}$ and $\omega_{\pi,0}$ both nonzero)
no general result exists which proves such an
assumption. However, with the additional assumption that, in an infinite
lattice, the breather
amplitude (measured at the level of its largest oscillation) can be lowered to
arbitrary small values \cite{Fla96,Flach97}, we come to the conclusion that, in
this limit, the breather frequency tends to an edge 
of the phonon band and merges with the corresponding mode.
Indeed, for small amplitudes, the motion enters a
quasi-linear regime and then approaches some phonon mode. But in the same
time, the breather family (parametrised either by its amplitude, 
its frequency or its energy) has to lie outside the phonon band to avoid a 
resonance which would lead to its disappearance. And this is possible by
approaching an edge of the phonon band only. Notice however that 
one can find systems with breather families which do not possess
any small amplitude limit (see e.g. 
\cite{Kas04}). 
Such breather families are then not related with the
instabilities discussed above.

\subsection{Energy thresholds for discrete breathers} \label{ssVA}

We now turn to an important implication of our results on the possible
existence of energy thresholds for families of discrete 
breathers bifurcating tangently from band edge modes. It has already been
noticed in many places that such a nonzero activation energy for
discrete breathers is of practical relevance \cite{Flach97,Kas04} as it
surely affects their experimental detection and presumably their
contribution to 
thermodynamical properties of 
lattices
(see for example \cite{Eleftheriou03} for some work in this
direction). 

Let us consider a Hamiltonian lattice of $N$ coupled 
oscillators described by \eqref{Ham} with 
families of periodic orbits parametrised by their amplitude. 
Let us follow the periodic orbit corresponding to a band edge
mode as its amplitude increases from zero. As we have seen in section \ref{Sec4}, under
certain conditions, at a finite critical amplitude (or energy density) 
this mode will become unstable and bifurcate tangently. A family of
discrete breathers emerging from this bifurcation is, right at this point, 
identical to the mode from which it stems and therefore has the same energy.
Increasing the amplitude further and following this new orbit, we obtain its
energy as a function of its amplitude. It is worth noticing immediately that
in finite systems $N <\infty$, discrete breathers arising in this way 
exist above a certain energy threshold only. Indeed, due to their finite 
amplitude their energy is surely nonzero. 
The question is as whether this 
threshold persists in the thermodynamic limit, $N \rightarrow \infty$.

We then turn to the determination of the (total) bifurcation energy $E_q^{(n)}$
of a BEM partially isochronous up to order $n$, in the thermodynamic limit. 
According to \eqref{nrjbifinp} and \eqref{nrjbifinop}, the bifurcation energy
density for the two possible modes $q=0$ or $q=\pi$ can be cast into the
general form
\begin{equation} \label{genbifnrg}
\varepsilon^{(n)}_q =
\left(\frac{\omega_{q,0}^2-\omega_{q_c,0}^2}{2n\gamma_{q,n}}\right)^{1/n}\, ,
\end{equation}
where the coefficient $\gamma_{q,n}$ is defined by the low-energy behaviour of
the BEM frequency
\begin{equation} \label{genfreqBEM}
\omega_q^2=\omega_{q,0}^2+\gamma_{q,n}\varepsilon^n + o(\varepsilon^n)\, ,
\end{equation}
and where the wave number $q_c$ is the closest to $q$ ( 
$q_c=2\pi/N$ if $q=0$ and $q_c=\pi-2\pi/N$ if $q=\pi$). In any case,
$\omega_{q,0}^2-\omega_{q_c,0}^2 \sim N^{-2}$. We then find that
\begin{equation} \label{totbifenrg}
E_q^{(n)} = N \varepsilon^{(n)}_q \sim N^{1-\frac{2}{n}}\ \ (N \rightarrow
\infty)\, .
\end{equation}
\noindent
This means that if the BEM is not isochronous, ($n=1$), its total bifurcation 
energy vanishes as the lattice becomes infinite 
and so does the breather energy in this limit. 
No energy threshold exists in this case as already mentioned in
\cite{Fla96}. However, as soon as the
band edge mode bears some degree of isochronism, ($n>1$), its total bifurcation
energy either converges to a finite value ($n=2$) or simply diverges
($n>2$) and energy thresholds are thus expected. 

We note incidentally that expression \eqref{totbifenrg}, although valid for a
one-dimensional chain, bears some striking resemblance with its
multi-dimensional counterpart in the {\em non isochronous} case which reads
$E_q^{(d)} \sim N^{1-\frac{2}{d}}$ where $d$ is the dimension of the lattice
\cite{Fla96,Flach97}. Combination of both isochronism ($n$) and
dimensionality ($d$) leads immediately to the conlusion that the 
total bifurcation energy of a BEM scales like
\begin{equation} \label{totbifenrgisodim}
E^{(n,d)} \sim N^{1-\frac{2}{nd}}\ \ (N \rightarrow \infty)\, .
\end{equation}
Energy thresholds for discrete breather families bifurcating tangently 
from BEMs are thus expected as soon as one of the positive integers 
$n$ or $d$ is strictly greater than one.

As already mentioned in \ref{Bifenipsec}, the general expression
\eqref{genbifnrg} still holds in case the analyticity of the onsite potential 
is relaxed. Recent results obtained by Kastner in \cite{Kas04,Kaslong04}
prove for example that, for (1D) onsite potentials of the form $V(x) = x^2/2 +
|x|^r/r+o(|x|^r)$, where $r$ is any {\em real} number greater than 2,
\eqref{genbifnrg} and \eqref{genfreqBEM} are still valid and that $n=r/2-1$.
This way, the exponent $n$ governing the low-energy behaviour of the frequency
\eqref{genfreqBEM} can be tuned continuously over the whole range of positive
real numbers and may counterbalance the effect due to the dimensionality $d$.
This {\em nonintegral} partial isochronism opens up the possibility to satisfy
the inequality obtained from \eqref{totbifenrgisodim}
\begin{equation} \label{ineqisodim}
nd < 2,\ \ n \in \R^+\, ,\ d \in \N^+\, ,
\end{equation}       
ensuring the absence of energy threshold in dimension $d$, even in two-
or three-dimensional systems. 

Finally, we also mention the existence of energy thresholds for breathers in 
one-dimensional systems with
algebraically decaying long range interactions \cite{SFlach98}.
We can then identify three lattice properties which
lead to appearance of breather energy thresholds - dimensionality,
interaction range, and (partial) isochronism.
 
\subsection{Energy thresholds for discrete breathers revisited} \label{ssVB}

\subsubsection{DNLS equations for the slow modulations of BEMs} \label{sssVB1}

To conclude this paper, we propose to revisit 
certain results of the previous section by deriving a discrete nonlinear
Schr\"odinger equation (DNLS) 
for the slow modulations of small-amplitude partially
isochronous BEMs. We shall do it with the help of a method based on their
nonlinear dispersion relation (see for example
\cite{Dodd82,Remoiss99,Scott99}) which renders its derivation almost
straightforward. We would like to mention that this result, obtained by a
somehow heuristic method, is also confirmed
by a more rigorous multiple-scale analysis at least for the first orders of
isochronism.  


 
The method we will use to derive the DNLS equation is not properly speaking a 
multiple-scale expansion for the motions $x_l(t)$ which would read
$x_l(t)=\sum_{j \geq 1} \mu^j F_{j,l}(t_0,t_1,\cdots)$ where $\mu \ll 1$ is
a small parameter and where 
the different timescales are given by $t_m = \mu^m t$ (see for example 
\cite{Remoiss86,Bang95,Konotop96,Giannou04}). But it represents a
similar approximation and has the advantage to be more transparent. 
It is based on the nonlinear dispersion relation obeyed by
a BEM isochronous up to order $n$.

We first note that in the harmonic limit, plane waves $x_{q;l}(t)
\propto e^{i(ql-\omega_{q,0}t)}$, where $\omega_{q,0}$ is given in 
\eqref{omega2q0}, are solutions of the 
linearised equations of motion \eqref{eqmotQlin}. 
Looking for slow modulations of the $q=0$ and $q=\pi$ 
modes, we write them
\begin{equation} \label{xltsm}
x_{q;l}(t) = A_l e^{i(ql-\omega_{q,0}t)} + {\rm c.c.} + o(\mu)
\end{equation}
where "c.c." stands for "complex conjugate", $q \in \{0,\pi\}$ and the
small amplitude $A_l = \mu \psi_l$. Here $\mu$ is some small parameter. 
 
The function $\psi_l$ is slowly varying in space and time. It is found 
typically from a multiple-scale analysis that, for non isochronous modes, it  
assumes the form $\psi_l \equiv \psi_l(\mu^2 t,\mu l)$ 
(see for example \cite{Giannou04}).
For modes isochronous up to order $n$, a similar analysis yields $\psi_l \equiv
\psi_l(\mu^{2n} t,\mu^n l)$. This is easily understood from the nonlinear
dispersion relation $\omega^2_q \simeq \omega^2_{q,0}+\gamma_{q,n}
\varepsilon^n$. 
For a slowly modulated plane wave to exist, it is known that
nonlinearity has to compensate the dispersion of the wave packet. Now, the
nonlinear term in the previous relation is, in leading order, proportional to
$\varepsilon^n \sim \mu^{2n}$. Therefore, the nonlinear correction to the
frequency introduces the natural long-time scale $\mu^{2n} t$. 
On another hand, the dispersion of the wave packet arises from a (spatial)
Laplacian. If $\mu^p l$ is the space scale, dispersion introduces a
correction of order $\mu^{2p}$. This correction is able to compensate
the nonlinear effects only if $p=n$, hence the scaling of $\psi_l$.

Now, in leading order (harmonic regime), the energy density of a BEM reads 
$\varepsilon_q = 2\omega^2_{q,0} |A|^2$, $q \in \{0,\pi\}$ (the BEM 
amplitude $A$ is constant so that it doesn't depend on $l$). Re-instating in
the nonlinear dispersion and expanding the latter around $q=0$ or $q=\pi$
yields
\begin{equation} \label{enlwq}
\omega_0 - \omega_{0,0} \simeq \frac{\phi_2}{\omega_{0,0}}\, [1-\cos(q)]
+\frac{\gamma_{0,n}}{2\omega_{0,0}} (2\omega^2_{0,0})^n |A|^{2n}\ \ \ 
{\rm
and}\ \ \ \omega_{\pi} - \omega_{\pi,0} \simeq -
\frac{\phi_2}{\omega_{\pi,0}}\, [1-\cos(q-\pi)]
+\frac{\gamma_{\pi,n}}{2\omega_{\pi,0}} (2\omega^2_{\pi,0})^n |A|^{2n}\, .
\end{equation}
As explained in \cite{Dodd82,Remoiss99,Scott99}, we can now use the dispersion
relations above to find the equations governing the envelope of the slowly
varying amplitude $A$. 
This merely amounts to 
replacing $(\omega_k - \omega_{k,0})$ by $i\partial_t$, $(q-k)$ 
by $i\partial_l$ (with $k=0,\pi$), $A$ by the slowly varying
amplitude $A_l$ and finally to let it act on the amplitude
$A_l$. We then obtain,
\begin{eqnarray} 
i\partial_t A_l &=& -
\frac{\phi_2}{2\omega_{0,0}}(A_{l+1}+A_{l-1}-2A_l) 
+\frac{\gamma_{0,n}}{2\omega_{0,0}} (2\omega^2_{0,0})^n |A_l|^{2n} A_l\,
, \ \ \ {\rm (in-phase)} \label{DNLS0} \\
i\partial_t A_l &=& 
\frac{\phi_2}{2\omega_{\pi,0}}(A_{l+1}+A_{l-1}-2A_l) 
+\frac{\gamma_{\pi,n}}{2\omega_{\pi,0}} (2\omega^2_{\pi,0})^n |A_l|^{2n} A_l\,
, \ \ \ {\rm (out-of-phase)} \, . \label{DNLSpi}
\end{eqnarray}    
   
Several remarks are in order at this stage. First, the above derivation is not
rigorous. Nevertheless, it makes the role of isochronism through
the nonlinear dispersion relation transparent  and allows us to derive the
DNLS equations 
for the modulation of the BEMs very easily. Second, we remark that for non
isochronous potentials ($n=1$), \eqref{DNLS0} is exactly the equation (6) 
derived by Kivshar in \cite{Kiv93} (see also \cite{Kiv94}, eq. (5)). 
In this respect,
\eqref{DNLS0} and \eqref{DNLSpi} are generalisations to higher
order of isochronism. We would like to
mention as well that, for $n=1$, \eqref{DNLS0} and \eqref{DNLSpi} have been
obtained by a rigorous multiple-scale expansion in \cite{Giannou04} (up to the
fact that the discrete spatial derivative is replaced by a continuous one).

Equations \eqref{DNLS0} and \eqref{DNLSpi} provide now a physical
motivation for studying DNLS with higher order nonlinearities. Indeed, the
nonlinear exponent is shown to be directly related to the isochronism of
the BEM under consideration. We notice finally that, \eqref{DNLS0} and
\eqref{DNLSpi} describe both the BEMs and their modulations and that, for this
reason they make it possible to study the bifurcation process which leads from
the plane wave to a breather (within the degree of approximation inherent to
their derivation).         
  
\subsubsection{Rederivation of discrete breather energy thresholds} \label{sssVB2}
   
Now, the condition for the stability of a plane wave for the cubic DNLS
equation 
on a periodic lattice has been derived by Carr and Eilbeck in \cite{Carr85}. 
Studying an equation of the type $i\dot{A}_l+\chi (A_{l+1}+A_{l-1}-2A_l) +
\lambda |A_l|^2 A_l$, with $A_{l+N}=A_l$, $N$ being the number of lattice
sites and $ {\cal N} = \sum_l |A_l|^2 = 1$, these authors show that the plane
wave $A_l = e^{i\omega t} A$ is stable if and only if $\lambda \leq 2N\chi
\sin^2 (\pi/N)$. The generalisation of this result to arbitrary nonlinearity
($\lambda |A_l|^{2n} A_l$) and norm ${\cal N}$ is straightforward and reads 
\begin{equation} \label{stabpwordn}
\frac{{\cal N}}{N} \leq \left[ \frac{2\chi}{n\lambda} \sin^2
(\pi/N)\right]^{1/n} \, .
\end{equation}
For a plane wave ${\cal N}/N = |A|^2$ and according to the mode $q$ under
consideration, the amplitude and the energy density are related via 
$\varepsilon_q = 2\omega_{q,0}^2 |A_q|^2$. 
Moreover, in \eqref{DNLS0} and \eqref{DNLSpi}, the
ratio of the coupling $\chi_q$ and the nonlinear parameter $\lambda_q$ is 
$\chi_q / \lambda_q = \pm \phi_2/ (2\omega_{q,0}^2)^n \gamma_{q,n}$, ``$+$'' for
$q=\pi$ and ``$-$'' for $q=0$. Then, eventually, we find the energies at
which the 
in-phase and out-of-phase modes become respectively unstable
\begin{equation} \label{stabnrgordn}
\varepsilon_{0}^{(n)} = \left(\frac{-2\phi_2
\sin^2\left(\frac{\pi}{N}\right)}{n\gamma_{0,n}}\right)^{1/n} \, ,
\ \ \ {\rm and}\ \ \ 
\varepsilon_{\pi}^{(n)} = \left(\frac{2\phi_2
\sin^2\left(\frac{\pi}{N}\right)}{n\gamma_{\pi,n}}\right)^{1/n} \, , 
\end{equation}
which are precisely the expressions obtained in the previous sections. Notice
that, these energies exist provided $\gamma_{0,n}<0$ and $\gamma_{\pi,n}>0$
respectively. At the level of \eqref{DNLS0} and \eqref{DNLSpi}, these
conditions 
imply $\chi_q  \lambda_q > 0$, ($q=0,\pi$), which is precisely the
condition for these equations to support bright type breathers.        

Our concluding remark is that, for nonlinearities higher than the usual cubic
one, the DNLS equation is known to possess energy thresholds for discrete
breathers (see for example \cite{Malomed96,Weinstein99,Flach97,Kevrekidis00}). 
This
allows us to corroborate our previous prediction that discrete breathers
steming from the tangent bifurcation of partially isochronous band edge modes
appear above certain energy thresholds only. Equations
\eqref{DNLS0} and \eqref{DNLSpi} should provide a way to estimate them.  
  
\section{Conclusions}

In this paper we have shown that, in network of coupled oscillators described
by Hamiltonian \eqref{Ham}, the way band edge modes (BEMs) possibly undergo a
tangent bifurcation crucially relies on the low-energy behaviour of their
frequency. This behaviour is obtained by expanding the frequency as a power
series in the energy (see \eqref{genfreqBEM}) as explained in section
\ref{sssIIB2} and in appendix \ref{App1} for the in-phase mode or in section
\ref{solople} for the out-of-phase mode.

The energy at which such modes may bifurcate tangently 
is given in
\eqref{nrjbifinp} for an in-phase mode and in \eqref{nrjbifinop} for an
out-of-phase mode. In these expressions, 
$n$ is the degree of isochronism of the mode under consideration and
$\gamma_{q,n}$, the first nonzero nonlinear coefficient of the low-energy
expansion of the frequency as defined in \eqref{genfreqBEM}. 

As proved in sections \ref{sssIVB7} and \ref{sssIVC6}, purely isochronous
modes (i.e. with an energy-independent frequency) never undergo a tangent
bifurcation. This result is {\em non perturbative}. Now, in a network
of coupled isochronous {\em onsite} potentials, the in-phase mode is
isochronous and never bifurcates tangently according to the previous theorem.
But the out-of-phase mode, which is generically non isochronous,
may bifurcate as soon as the nonlinearity of the {\em interaction}
potential is strong enough as expressed in \eqref{crline}. 

Bifurcation energies obtained in \eqref{nrjbifinp} and \eqref{nrjbifinop} are
used in section \ref{ssVA} to derive a condition for the occurrence of energy
thresholds for discrete breathers bifurcating tangently from BEMs. For modes
isochronous up to order $n$ and in dimension $d$, this reads $nd \geq 2$. For
analytic potentials ($n$ positive integer) energy thresholds exist whenever
$n\geq 2$ (even in 1D) or when $d\geq 2$. For non-analytic potentials, $n$
becomes a positive real number which opens up the possibility of absence of
energy threshold even in 2- or 3D.

In section \ref{sssVB1}, we derived two DNLS equations
respectively related to the slow modulations of small-amplitude
partially isochronous in-phase (eq. \eqref{DNLS0}) and out-of-phase
(eq. \eqref{DNLSpi}) modes. These equations are the generalisation to higher
degree of isochronism of the usual cubic DNLS equation. 
Their nonlinearity is proportional to the degree of isochronism of
the corresponding mode \footnote{Note that the in- and out-of-phase degrees
  are independent of each other. For a given system, equations \eqref{DNLS0}
  and \eqref{DNLSpi} will then generally have different nonlinearities.}.     
As shown in section \ref{sssVB2}, they allow for an easy rederivation of the
bifurcation energies \eqref{nrjbifinp} and \eqref{nrjbifinop}.

\section*{Acknowledgements}

It is a pleasure to thank M. Kastner, J.C. Eilbeck and W. Dreyer for
stimulating discussions and helpful comments. 
J. Dorignac would like to thank the Max-Planck-Institut f\"ur Physik komplexer
Systeme (Dresden, Germany) for supporting his research during the period of
his stay.  
  
\appendix
\section{Energy expansion for the period in a 1D potential} \label{App1}

We derive in this appendix a simple method to compute the period of
oscillation  
around the minimum of a potential $V(x)$ as a power series in
the energy $E$. This method is, to our opinion, simpler than other 
methods which have been developed earlier \cite{Rodriguez00,Fernandez01}.
It yields the coefficients of this power series in terms of the 
coefficients of the Taylor expansion of $V(x)$ around its minimum. 
This is used to construct a potential partially isochronous up to 
order $n$, which merely amounts to canceling out all the coefficients 
of the power series in $E$ up to order $n-1$.

We assume that $V(x)$ has the following properties: $V(0)=0, V'(0)=0,
V"(0)=\omega^2 \neq 0$. Moreover, we require that $V'(x)/x > 0$ on an interval
$I$ centered around the minimum. In this interval, 
$V(x)$ can be expanded as a power series in $x$ around 0, 
\begin{equation} \label{Vtayl}
V(x) = \frac{1}{2}\omega^2 x^2 + \omega^2 \sum_{n=3}^{\infty} \frac{1}{n}
\alpha_{n-1} x^n .
\end{equation} 
At a given energy $E$, the period $T$ is given by
\begin{equation} \label{per1}
T(E) = \sqrt{2} \int_{x_{-}(E)}^{x_{+}(E)} \frac{dx}{\sqrt{E-V(x)}}
\end{equation} 
where $x_{\pm}(E)$ are the two solutions of $V(x)=E$, ($x_{+}(E)>0$ and
$x_{-}(E)<0$). Let us make a change of variable defined by 
$V(x) = \frac{1}{2}\omega^2 X^2$, with $dX(x)/dx > 0$ for $x \in I$. The last
condition ensures that $X$ is a monotonic increasing function of $x$. 
Then (\ref{per1}) yields,
\begin{equation} \label{per2}
T(E) = \sqrt{2} \int_{0}^{\frac{\sqrt{2E}}{\omega}}
\frac{\Delta'(X)}{\sqrt{E-\frac{1}{2}\omega^2 X^2}}\, dX = \frac{2}{\omega}
\int_0^1 \Delta'\left(\frac{\sqrt{2E}}{\omega} u\right) \frac{du}{\sqrt{1-u^2}}
\end{equation}
where $\Delta(X)=x(X)-x(-X)$ with $dx(X)/dX > 0$. $\Delta'$ represents the
derivative with respect to $X$. In order to compute the last expression of
$T(E)$, we need to invert the change of variable. Taking into account the condition $dx(X)/dX > 0$, we obtain
\begin{equation} \label{xofX}
V(x) = \frac{1}{2}\omega^2 X^2 \Rightarrow x(X) = X +\sum_{n=2}^{\infty}
\sigma_n X^n 
\end{equation}
where the coefficients $\sigma_n$ are functions of the $\alpha_m$.
Hence
\begin{equation} \label{DofX}
\Delta'(X) = 2\left(1 +\sum_{n=1}^{\infty} (2n+1) \sigma_{2n+1} X^{2n}\right).
\end{equation}
Reinstating this expression into (\ref{per2}) and performing a term by term
integration over $u$, we finally get
\begin{equation} \label{per3}
T(E) = \frac{2\pi}{\omega}\left[ 1+\sum_{n=1}^{\infty}
\sigma_{2n+1}\frac{(2n+1)!!}{(2n)!!}\left(\frac{2E}{\omega^2}\right)^n \right]
\end{equation}
The inversion (\ref{xofX}) is easily done with the help of any software able
to perform formal calculations. Using Maple \cite{Maple}, we have computed the
few first odd coefficients $\sigma_{2n+1}$:
\begin{eqnarray} \label{sigman}
\sigma_3 &=& \frac{5}{18}\alpha_2^2-\frac{1}{4}\alpha_3 \nonumber \\
\sigma_5 &=& \frac{77}{216}\alpha_2^4 - 
\frac{7}{8}\alpha_3\alpha_2^2
 + \frac{7}{15}\alpha_4\alpha_2
 + \frac{7}{32}\alpha_3^2 - \frac {1}{6}\alpha_5 \nonumber \\
\sigma_7 &=& \frac {3}{7}\alpha _{6}\alpha _{2} - 
\frac {33}{20}\alpha _{4}\alpha _{3}
\alpha _{2} + \frac {143}{90}\alpha _{4}
\alpha _{2}^{3} + \frac {9}{50}\alpha _{4}^{2} - \frac {715}{288}\alpha
_{2}^4 \alpha _{3} + \frac {143}{64} \alpha _{3}^{2
}\alpha _{2}^{2} - \frac {11}{12} \alpha_{5}\alpha _{2}^{2} + \frac {3}{8}
\alpha _{3}\alpha_{5} + \frac {2431}{3888} \alpha _{2}^{6} - 
\frac {33}{128} \alpha _{3}^{3} - 
\frac {1}{8} \alpha _{7} \nonumber 
\end{eqnarray}  
To construct a potential partially isochronous up to order $n$ we need to
zero the coefficients $\sigma_{2k+1}$ up to $\sigma_{2n-1}$. This
constraints the coefficients $\alpha_k$ to verify certain relations 
given below:
\begin{eqnarray}
\text{Order} \ n=2 \ \ \ &;&\ \ \ \alpha_3 = \frac{10}{9}\alpha_2^2 \nonumber \\
\text{Order} \ n=3 \ \ \ &;&\ \ \ \alpha_5 = -\frac{56}{27}\alpha_2^4 + 
\frac{14}{5}\alpha_4 \alpha_2 \nonumber \\
 \text{Order} \ n=4 \ \ \ &;&\ \ \ \alpha_7 = \frac{24}{7}\alpha_6 \alpha_2  
-\frac{592}{45}\alpha_4 \alpha_2^3 + \frac{36}{25}\alpha_4^2 
+ \frac{848}{81}\alpha_2^6 \nonumber 
\end{eqnarray} 

\section{Solvability condition for system (\ref{Eqrj})} \label{App2}

We prove hereafter that, up to order $2n$ in the amplitude $y$, system
\eqref{Eqrj} never develops any secular term.  
\begin{itemize}
\item Up to order $y^{2n}$, we have shown in section \ref{secIVC3} that 
the real parts $r_{\pi}$ and $r_{0}$ of
the two modes $\eta_{\pi-2\pi/N}$ and $\eta_{2\pi/N}$ 
obey the system
\begin{equation}
\left\{
\begin{array}{ccc}
\displaystyle{\ddot{r}_{\pi}+{\cal T}_1(0,\tau)r_{\pi}+{\cal T}_2(\tau)r_{0}} & = & 0,\\
& & \\
\displaystyle{\ddot{r}_{0}+{\cal T}_4(0,\tau)r_{0}+{\cal T}_2(\tau)r_{\pi}} & = & 0,
\end{array}
\right.
\end{equation}
The functions ${\cal T}_{i}(\tau)$ are $2\pi$-periodic.
\item Up to the same order, the two periodic solutions of this system are
given by
\begin{equation}
\left(
\begin{array}{c}
r_{\pi}^{(s)} \\
\\
r_{0}^{(s)}
\end{array} \right) = \left(
\begin{array}{c}
\displaystyle{\frac{\partial Q_{\pi}}{\partial y}}\\
\\
\displaystyle{\frac{\partial Q_{0}}{\partial y}}
\end{array} \right) \ \ \ \text{and}\ \ \ 
\left(
\begin{array}{c}
r_{\pi}^{(a)}\\
\\
r_{0}^{(a)}
\end{array} \right) = \left(
\begin{array}{c}
\displaystyle{\frac{1}{y}\frac{\partial Q_{\pi}}{\partial \tau}}\\
\\
\displaystyle{\frac{1}{y}\frac{\partial Q_{0}}{\partial \tau}}
\end{array} \right)
\end{equation}
The first solution is time-symmetric (${}^{(s)}$) whereas
the second is time-antisymmetric (${}^{(a)}$).
\item From order 0 to $n$ we can take the imaginary parts $j_{\pi}$ and
$j_{0}$ of the modes equal to 0. From $(n+1)$ to $2n$ they obey,
\begin{equation} \label{Eqj}
\left\{
\begin{array}{ccc}
\displaystyle{\ddot{j}_{\pi}+{\cal T}_1(0,\tau)j_{\pi}+{\cal T}_2(\tau)j_{0}+{\cal T}_3(\delta,\tau)r_{0}^{(s)}} &
= & 0,\\ 
& & \\
\displaystyle{\ddot{j}_{0}+{\cal T}_4(0,\tau)j_{0}+{\cal T}_2(\tau)j_{\pi}-{\cal T}_3(\delta,\tau)r_{\pi}^{(s)}} & = & 0,
\end{array}
\right.
\end{equation}
where ${\cal T}_3$ is time-periodic and time-symmetric (it is a function of
$Q_{\pi}$ which is time-symmetric itself). 

\item The problem is to prove that from $y^{(n+1)}$ to
$y^{2n}$, (\ref{Eqj}) never develops secular terms. We will use the
solvability condition below (see for example \cite{Farkas94}).

\end{itemize} 

\begin{thm}[Fredholm alternative]
Assume the subspace of $T$-periodic solutions of the homogeneous system
$$ \frac{d}{dt} \ke{x} = A(t) \ke{x} ,$$ 
$A(t+T)=A(t)$, $A \in {\rm Mat}(n \times n)$ is of dimension $k\geq 1$. Denote by $\ke{\eta^{(l)}} ,\ l \in \{1,\cdots ,k\}$,
$k$ linearly independent $T$-periodic solutions of the adjoint system 
$$ \frac{d}{dt} \ke{\eta} = -A^{\dagger}(t) \ke{\eta} . $$
For $\ke{f(t+T)}=\ke{f(t)}$, the inhomogeneous system
$$ \frac{d}{dt} \ke{y} = A(t) \ke{y} +\ke{f(t)},$$
has $T$-periodic solutions if and only if 
\begin{equation} \label{solvcond}
\int_0^T \bk{\eta^{(l)}}{f(t)}\, dt = 0 \ \ \ (l \in \{1,\cdots ,k\}) .
\end{equation}
\end{thm}

\begin{itemize}
\item In our case, we rewrite (\ref{Eqj}) as a first order inhomogeneous
system with $\ke{y} = {\rm col}(r_{\pi},r_0,p_{\pi},p_0)$ (column vector) and 
$p_{\pi}=\dot{r}_{\pi},p_0=\dot{r}_0$. We have 
\begin{equation}
A = \left( \begin{array}{cccc}
0 & 0 & 1 & 0 \\
0 & 0 & 0 & 1 \\
-{\cal T}_1 & -{\cal T}_2 & 0 & 0 \\
-{\cal T}_2 & -{\cal T}_4 & 0 & 0 
\end{array} \right) \ \ \ \text{and}\ \ \ 
\ke{f} = \left( \begin{array}{c}
0 \\
0 \\
-{\cal T}_3 r_0^{(s)}\\
{\cal T}_3 r_{\pi}^{(s)} 
\end{array} \right)
\end{equation}
The corresponding homogeneous adjoint system is 
\begin{equation}
\frac{d}{dt} \ke{\eta} = -A^{\dagger}(t) \ke{\eta} \Leftrightarrow 
\frac{d}{dt} \left( \begin{array}{c}
\eta_1 \\
\eta_2 \\
\eta_3 \\
\eta_4 
\end{array} \right) = \left( \begin{array}{cccc}
0 & 0 & {\cal T}_1 & {\cal T}_2  \\
0 & 0 & {\cal T}_2 & {\cal T}_4  \\
-1 & 0 & 0 & 0 \\
0 & -1 & 0 & 0 
\end{array} \right) \left( \begin{array}{c}
\eta_1 \\
\eta_2 \\
\eta_3 \\
\eta_4 
\end{array} \right)  
\end{equation}
Its solutions are
\begin{equation}
\left( \begin{array}{c}
\eta_1 \\
\eta_2 \\
\eta_3 \\
\eta_4 
\end{array} \right)^{(s),(a)} =  \left(
\begin{array}{c}
-p_{\pi}\\
-p_{0}\\
r_{\pi}\\
r_{0}
\end{array} \right)^{(s),(a)}
\end{equation}

\item Calculating the two solvability conditions (\ref{solvcond}), we get
\begin{equation}
\int_0^{2\pi} \bk{\eta^{(s)}}{f}\, dt = \int_0^{2\pi}
{\cal T}_3 (-r_{\pi}^{(s)} r_0^{(s)} + r_0^{(s)} r_{\pi}^{(s)})\, d\tau = 0
\end{equation}
and
\begin{equation}
\int_0^{2\pi} \bk{\eta^{(s)}}{f}\, dt = \int_0^{2\pi}
{\cal T}_3 (-r_{\pi}^{(a)} r_0^{(s)} + r_0^{(a)} r_{\pi}^{(s)})\, d\tau = 0
\end{equation}
Indeed, to compute this last expression, we remember that ${\cal T}_3$ is
time-symmetric (series of cosine terms). The term in parenthesis being
time-antisymmetric, the integrand is 
time-antisymmetric. Because it is $2\pi$-periodic, the integral is zero.  

\item Conclusion: The solvability conditions are fulfilled and consequently
(\ref{Eqj}) does not develop any secular terms up to order $2n$. The way we
have obtained the critical energy for the tangent bifurcation is thus correct.
   
\end{itemize}


\begin{thebibliography}{9}

\bibitem{Sievers95}
A. J. Sievers and J. B. Page, in:
{\em Dynamical properties of solids VII phonon physics the cutting edge},
Elsevier, Amsterdam (1995).

\bibitem{Aubry97} 
S. Aubry, 
{\em Physica D} {\bf 103}, 201 (1997).


\bibitem{Flachrep98}
S. Flach and C.R. Willis, 
{\em Phys. Rep.} {\bf 295}, 181 (1998).


\bibitem{MacKay00}
R.S. MacKay, 
{\em Physica A} {\bf 288}, 174 (2000).

\bibitem{FlachZol01}
S. Flach and Y. Zolotaryuk, 
{\em Adv. in Solid State Phys.} {\bf 41}, 315 (2001).

\bibitem{Dauxois04}
T. Dauxois, A. Litvak-Hinenzon, R. S. MacKay and A. Spanoudaki (Eds.),
Energy Localisation and Transfer, Advanced Series in Nonlinear Dynamics
{\bf 22}, World Scientific Singapore (2004). 

\bibitem{Campbell04}
D. K. Campbell, S. Flach and Yu. S. Kivshar,
{\em Physics Today}, p.43, January (2004).

\bibitem{MacKay94}
R.S. MacKay and S. Aubry, 
{\em Nonlinearity} {\bf 7}, 1623 (1994).

\bibitem{Sepulchre97}
J.A. Sepulchre and R.S. MacKay,
{\em Nonlinearity} {\bf 10}, 679 (1997).

\bibitem{Flach95}
S. Flach, 
{\em Phys. Rev. E} {\bf 51}, 1503 (1995).

\bibitem{James01}
G. James, 
{\em C. R. Acad. Sci. Paris} {\bf 332}, S\'erie I, 581 (2001).

\bibitem{Aubry01}
S. Aubry, G. Kopidakis and V. Kadelburg, 
{\em DCDS-B} {\bf 1}, 271 (2001).

\bibitem{Fla96}
S. Flach, 
{\em Physica D} {\bf 91}, 223 (1996).

\bibitem{bb83}
N. Budinsky and T. Bountis,
{\em Physica D} {\bf 8}, 445 (1983).


\bibitem{Flach97}
S. Flach, K. Kladko and R.S. MacKay,
{\em Phys. Rev. Lett.} {\bf 78} (7), 1207 (1997).

\bibitem{Weinstein99}
M.I. Weinstein,
{\em Nonlinearity} {\bf 12}, 673 (1999).

\bibitem{Bolotin}
S.V. Bolotin, R.S. MacKay, 
in ``Localisation and energy transfer in nonlinear
systems'', eds L. Vazquez, R. S. MacKay and
M. P. Zorzano, World Scientific Singapore,
p.217 (2003).

\bibitem{Rodriguez00}
I. Rodriguez and J.L. Brun,
{\em Eur. J. Phys.} {\bf 21}, 617 (2000).

\bibitem{Fernandez01}
F.M. Fern\'andez,
{\em Eur. J. Phys.} {\bf 22}, 639 (2001).

\bibitem{Maple}
Maplesoft, a division of Waterloo Maple Inc. 2004, http://www.maplesoft.com

\bibitem{Nayfey}
A.H. Nayfeh, Introduction to Perturbation Techniques, Wiley Classics Library
Edition, 1993, Chapter 11.

\bibitem{Osypowski}
E.T. Osypowski and M.G. Olsson,
{\em Am. J. Phys.} {\bf 55} (8), 720 (1987).

\bibitem{Kas04}
M. Kastner,
{\em Phys. Rev. Lett.} {\bf 92} (10), 104301 (2004).

\bibitem{Kaslong04}
M. Kastner,
{\em arXiv:nlin 0401038} (2004). 

\bibitem{Magnus79}
W. Magnus and S. Winkler, Hill's equation, chap. VII, Dover ed. (1979).

\bibitem{Rand04}
R.H.Rand, Lecture Notes on Nonlinear Vibrations,
Published on-line by The Internet-First University Press (2004),
http://dspace.library.cornell.edu/handle/1813/79

\bibitem{Nieto81}
M.M. Nieto a,d V.P. Gutschick,
{\em Phys. Rev. D} {\bf 23} (4), 922 (1981).

\bibitem{Eleftheriou03}
M. Eleftheriou, S. Flach and G. P. Tsironis,
{\em Physica D} {\bf 186}, 20 (2003);
M. Ivanchenko, O. Kanakov, V. Shalfeev and S. Flach,
{\em Physica D} {\bf 198}, 120 (2004);
M. Eleftheriou and S. Flach, cond-mat/0409450 (2004).        

\bibitem{FlachDorignac03}
S. Flach, J. Dorignac, A. E. Miroshnichenko and V. Fleurov,
{\em Int. J. Mod. Phys. B} {\bf 17}, 3996 (2003). 

\bibitem{SFlach98}
S. Flach,
{\em Phys. Rev. E} {\bf 58}, R4116 (1998). 

\bibitem{Dodd82}
R.K Dodd, J.C. Eilbeck, J.D. Gibbon, H.C. Morris,
Solitons and Nonlinear Wave Equations, Academic press, 1982, p. 497.

\bibitem{Remoiss99}
M. Remoissenet,
Waves Called Solitons, Third edition, Springer-Verlag, 1999, p. 73.       

\bibitem{Scott99}
A.C. Scott,
Nonlinear Science, Emergence \& Dynamics of Coherent Structures, 
Oxford applied and engineering mathematics, 1999, p. 94. 

\bibitem{Remoiss86}
M. Remoissenet,
{\em Phys. Rev. B} {\bf 33} (4), 2386 (1986).

\bibitem{Bang95}
O. Bang and M. Peyrard,
{\em Physica D} {\bf 81}, 9 (1995).

\bibitem{Konotop96}
V.V. Konotop,
{\em Phys. Rev. E} {\bf 53}, 2843 (1996).

\bibitem{Giannou04}
J. Giannoulis and A. Mielke,
{\em Nonlinearity} {\bf 17}, 551 (2004).

\bibitem{Kiv93}
Y.S. Kivshar,
{\em Phys. Lett. A} {\bf 173}, 172 (1993).

\bibitem{Kiv94}
Y.S. Kivshar, M. Haelterman and A.P. Sheppard,
{\em Phys. Rev. E} {\bf 50} (4), 3161 (1994).
 
\bibitem{Carr85}
J. Carr and J.C. Eilbeck,
{\em Phys. Lett. A} {\bf 109}, 201 (1985).

\bibitem{Malomed96}
B. Malomed and M.I. Weinstein,
{\em Phys. Lett. A} {\bf 220}, 91 (1996). 

\bibitem{Kevrekidis00}
P.G. Kevrekidis, K.O. Rasmussen and A.R. Bishop,
{\em Phys. Rev. E} {\bf 61} (4), 4652 (2000).

\bibitem{Farkas94}
M. Farkas, Periodic motions, Applied Mathematical Sciences 104,
Springer-Verlag, 1994, p. 62. 

\end{thebibliography}
\end{document}